%% file: v3_TWC_I.tex
\documentclass[letterpaper,journal]{IEEEtran}

\usepackage{amsmath,amssymb,amsthm,mathtools,bm,bbm}
\usepackage{mathrsfs}
\usepackage{cases}
\usepackage{braket}

\usepackage{algorithm}
\usepackage{algorithmic}

\usepackage{graphicx}
\usepackage{booktabs}
\usepackage{multirow}
\usepackage{threeparttable}
\usepackage{siunitx}
\usepackage{tablefootnote}

\usepackage{tikz}
\usetikzlibrary{arrows.meta,calc,positioning,decorations.pathreplacing,patterns,fit}
\usepackage{pgfplots}
\pgfplotsset{compat=1.18}
\usetikzlibrary{plotmarks}

\usepackage{xcolor}
\usepackage{balance}
\usepackage{eqparbox}
\usepackage{setspace}
\usepackage{verbatim}
\usepackage{dirtytalk}

\usepackage{cite}
\usepackage{hyperref}
\newtheorem{lemma}{Lemma}

\newcommand{\R}{\mathbb{R}}
\newcommand{\C}{\mathbb{C}}

\newcommand{\I}{\mathbf{I}}
\newcommand{\CN}{\mathcal{CN}}
\newcommand{\vecc}{\operatorname{vec}}
\newcommand{\Hsf}{\mathsf{H}}

\DeclareMathOperator{\rank}{rank}
\DeclareMathOperator{\tr}{tr}
\DeclareMathOperator{\diag}{diag}
\DeclareMathOperator{\sinc}{sinc}

\usepackage{todonotes}

\newcommand{\orcidauthorA}{\href{https://orcid.org/0000-0001-5792-0842}{\includegraphics[scale=0.05]{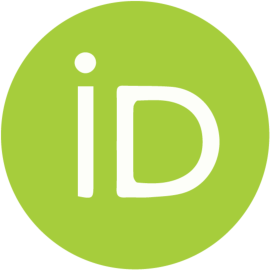}}}
\newcommand{\orcidauthorB}{\href{https://orcid.org/0000-0002-1478-2272}{\includegraphics[scale=0.05]{Figures/orcid_16x16.eps}}}

\begin{document}

%=== TITLE & AUTHORS ====================================================================
\title{A Novel Geometry-Aware GPR-Based Energy-Efficient and Low-Overhead Channel Estimation Scheme}
	\author{{Syed Luqman Shah$^{\orcidauthorA{}}$, \textit{Graduate Student Member}, \textit{IEEE} and Nurul Huda Mahmood$^{\orcidauthorB{}}$, \textit{Member}, \textit{IEEE}
    }\\
    \thanks{The authors S. L. Shah and N. H. Mahmood are with Centre for Wireless Communications, University of Oulu, Finland (e-mail: \{syed.luqman, nurulhuda.mahmood\}@oulu.fi.)\\
    \indent This work was supported by the Research Council of Finland (359850 6G-ConCoRSe, and 369116 \href{https://www.6gflagship.com/}{6G Flagship}).}}

%====================================================================
\maketitle
\begin{abstract}
Accurate channel state information (CSI) acquisition under tight pilot and training-energy budgets is crucial for next-generation wireless networks. We reduce pilot overhead and training energy by activating only a subset of transmit antennas during training. This yields sparse and noisy observations of the channel coefficients, casting CSI acquisition as a highly underdetermined Bayesian inverse problem. To solve this problem, we model the wireless channel as a proper complex Gaussian process (GP) over the joint transmit--receive antenna-index lattice and develop a GPR framework that reconstructs the full CSI by extrapolating observed pairs to unobserved ones. The GP prior is specified through a novel array-geometry-based spectral-mixture covariance function (GB-SMCF), which is Hermitian positive-semidefinite. Unlike standard distance-based kernels, the proposed GB-SMCF captures both smooth spatial decay and the oscillatory correlations induced by multipath propagation. The hyperparameters of the GB-SMCF are learned online within each coherence block by maximizing the marginal likelihood of the sparse pilot observations, requiring neither offline training nor prior knowledge of the channel covariance. Numerical results show that the proposed estimator reduces pilot overhead by up to $75\%$ and total training energy by up to $93.75\%$, while achieving lower normalized mean-square error than the considered baselines. It also yields higher spectral efficiency in the low-to-moderate SNR regime.
\end{abstract}

%%%%%%%%%%%%%%%%%%%%%%%%%%%%%%%%%%%%%%%%%%%%%%%%%%%%%%%%%%%%%%%%%%%%%%%%%%%%%%%%%%%%%%%%%%

%%%%%%%%%%%%%%%%%%%%%%%%%%%%%%%%%%%%%%%%%%%%%%%%%%%%%%%%%%%%%%%%%%%%%%%%%%%%%%%%%%%%%%%%%%
%%%%%%%%%%%%%%%%%%%%%%%%%%%%%%%%%%%%%%%%%%%%%%%%%%%%%%%%%%%%%%%%%%%%%%%%%%%%%%%%%%%%%%%%%%
\vspace{-2mm}
\section{Introduction}
Multiple-input multiple-output (MIMO) is a key enabler of sixth-generation (6G) wireless systems, providing spatial multiplexing and beamforming gains~\cite{Nandana_white, R1_stackedRIS}. However, realizing these gains requires accurate acquisition of channel state information (CSI)~\cite{WeiBer2026}, which is challenging particularly in local 6G deployments such as compact indoor cells~\cite{shah2025,ShahD2D}, where rich scattering, large apertures, and wavefront curvature induce structured spatial correlations across the array~\cite{Li2024NearField}. Measurements and electromagnetic modeling show that these correlations are governed by array geometry, scattering clusters, and angular characteristics~\cite{Jieao2025IEEE_TIT, Emil_Correlation_2024_Conference}. For uniform rectangular arrays (URAs), they appear as slowly varying spatial envelopes modulated by oscillatory components associated with dominant angles of arrival and departure~\cite{Emil_Correlation_2024_Conference}. This structure provides a natural basis for covariance-aware channel estimation~\cite{WeiBer2026, li2024spatioGPR}. Gaussian process regression (GPR) provides a Bayesian mechanism for exploiting such correlations because its covariance function (kernel) encodes the second-order structure for posterior prediction~\cite{rasmussen2006gaussian}. When the kernel is designed to reflect antenna geometry and propagation physics~\cite{Jieao2025IEEE_TIT}, GPR can reconstruct unobserved channel coefficients from sparse pilot observations without relying on explicit sparsity, bandlimitedness, or fixed covariance assumptions~\cite{shah2026improved}. Motivated by this structure, we review pilot-based CSI acquisition methods and identify the gap addressed herein.

%%%%%%%%%%%%%%%%%%%%%%%%%%%%%%%%%%%%%%%%%%%%%%%%%%%%%%%%%%%%%%%%%%%%%%%%%%%%%%%%%%%%%%%%%%
\vspace{-3mm}
\subsection{Related Works}
Pilot-aided training is the standard approach for CSI acquisition, wherein known reference sequences are sounded over the channel to infer the propagation response from the received observations~\cite{fodor2017overview}. Under orthogonal training, the least-squares (LS) and minimum mean-square error (MMSE) estimators require a pilot length at least equal to the number of transmit antennas~\cite{Emil2024}, and this requirement scales unfavorably with increasing antenna counts~\cite{shah2025lowoverhead, shah2026improved}.

Compressed sensing (CS)-based methods reduce pilot overhead by exploiting angular or delay sparsity~\cite{Tony_OMP_TIT2011}. Dictionary-based algorithms, such as orthogonal matching pursuit (OMP) and approximate message passing (AMP), recover the channel from fewer pilots when the propagation environment is dominated by a small number of grid-aligned specular paths~\cite{Tony_OMP_TIT2011, donoho2010message}. However, continuous angular spreads, near-field propagation, and rich scattering can violate the discrete sparsity model, leading to basis mismatch, power leakage, and reconstruction bias~\cite{ADMM_algo_Sig_letter_2018}. Matrix-completion methods offer an alternative means of recovering missing entries by assuming an exploitable low-rank structure and solving a nuclear-norm relaxation problem~\cite{Matrix_Completion_Zhiwei2025}. These methods are effective when the dominant spatial structure is globally coherent across the array, but strong local variations and aperture-dependent propagation effects can weaken the low-rank assumption~\cite{Li2024NearField}.

Gaussian process (GP)-based models provide a principled Bayesian alternative for CSI acquisition by combining statistical regularization with covariance-based prior modeling~\cite{WeiBer2026}. Existing GP and Gaussian-prior approaches differ mainly in how they encode channel structure. Distance-based GPR methods model the channel as a Gaussian field using kernels such as the squared-exponential (SqExp) and Mat\'ern kernels, enabling full-CSI prediction from partial observations with reduced pilot overhead~\cite{shah2025lowoverhead,shah2026improved}. Physics-informed GP priors further derive covariance functions from electromagnetic principles, thereby incorporating angular dispersion and near-field effects into MMSE estimation and channel prediction~\cite{li2024spatioGPR,Jieao2025IEEE_TIT}. Beyond kernel-based GP modeling, conditionally Gaussian priors have been used in semi-blind estimation to exploit both pilot and payload observations through expectation-maximization and MMSE-type inference~\cite{WeiBer2026,Nayebi2018}. Hierarchical and pattern-coupled Gaussian priors have been used in sparse Bayesian learning to impose structured sparsity, common support, and robustness to outliers or hardware-induced perturbations~\cite{Cheng2018,xiao2021channel,Arj25}. Learned Bayesian estimators further use Gaussian-mixture or generative priors to approximate non-Gaussian channel distributions in beamspace and quantized MIMO estimation~\cite{Wei2021deep,Fesl2024enhancing}. Beyond CSI acquisition, GP-based models have also been used for interference prediction in local 6G networks~\cite{shah2025}.

%%%%%%%%%%%%%%%%%%%%%%%%%%%%%%%%%%%%%%%%%%%%%%%%%%%%%%%%%%%%%%%%%%%%%%%%%%%%%%%%%%%%%%%%%%
\vspace{-2mm}
\subsection{Motivation and Contributions}
Despite these advances, existing GP-based and Gaussian-prior channel-estimation frameworks exhibit several key limitations. \textit{First}, their observation models are developed for full-array pilot sounding, semi-blind payload-assisted reception, beamspace projections, quantized measurements, or hardware-impaired observations~\cite{WeiBer2026,Nayebi2018,Wei2021deep,Fesl2024enhancing,xiao2021channel,Arj25}, rather than recovering unsounded channel coefficients from sparse antenna-indexed pilot samples. \textit{Second}, their structural priors encode channel structure in transformed or latent domains rather than directly over the physical antenna-index domain; sparse Bayesian learning imposes angular-domain sparsity through hierarchical or pattern-coupled priors~\cite{Cheng2018,xiao2021channel}, whereas learned and generative estimators rely on Gaussian-mixture or latent priors tailored to semi-blind, beamspace, or quantized inference models~\cite{WeiBer2026,Wei2021deep,Fesl2024enhancing}. \textit{Third}, although physics-informed GP with electromagnetic kernels derived from Maxwell’s equations over an angular distribution has improved covariance-aware estimation or channel-forecasting accuracy~\cite{li2024spatioGPR,Jieao2025IEEE_TIT}, it has not been used to enable reduced-training CSI completion. \textit{Fourth}, the required prior information is typically obtained externally; MMSE estimation assumes known or separately estimated channel covariance, whereas learned Bayesian estimators often rely on offline datasets or pretrained generative models~\cite{WeiBer2026,Wei2021deep,Fesl2024enhancing}. Consequently, reduced-pilot CSI completion from sparse antenna-indexed observations remains insufficiently explored.

To address these gaps, this paper develops a GPR-based CSI acquisition framework that combines subset-activated transmit training with an online-learned, geometry-aware, proper-complex GP prior over the joint transmit--receive antenna-index lattice to reduce pilot overhead and training energy while improving estimation accuracy and spectral efficiency (SE). Our key contributions are summarized as follows:

\begin{itemize}
    \item To reduce pilot overhead and training energy, we propose a training scheme in which only a subset of transmit antennas is activated to send orthogonal pilots, while all receive antennas listen. The resulting observations constitute a column-subsampled noisy measurement of the channel matrix. This yields an underdetermined linear inverse problem, which we formulate as a Bayesian reconstruction task.

    \item We model the wireless channel as a proper-complex GP over the joint transmit--receive antenna indices and develop a GPR-based estimation framework. This framework solves the Bayesian inverse problem through probabilistic interpolation over the joint transmit--receive antenna-index lattice and extrapolation to unobserved channel entries.

    \item To encode physically meaningful propagation structure and improve reconstruction accuracy, we develop an array-geometry-based spectral-mixture covariance function (GB-SMCF) and prove its validity as a Hermitian positive-semidefinite (PSD) kernel. The proposed GB-SMCF is a separable complex covariance function over the transmit and receive lattices that captures multiple angular components, smooth spatial envelopes, and oscillatory behavior through an interpretable spectral-mixture parameterization.

    \item The proposed GB-SMCF-based GPR learns all model hyperparameters online by maximizing the marginal likelihood of the current pilot observations. We further derive closed-form gradients, enabling efficient gradient-based optimization in each coherence block without offline training or external channel datasets. This yields closed-form posterior mean and covariance expressions for the full channel matrix, providing a statistically principled reconstruction of both observed and unobserved antenna pairs.
\end{itemize}

Collectively, these contributions establish a new paradigm: \textit{modeling a structured array-geometry-aware complex channel covariance function, learning it from sparse pilot observations, and using it for statistically principled CSI completion}. The method replaces the \say{known covariance} assumption of MMSE and the \say{known dictionary} assumption of CS with a learned, physically interpretable array-domain prior optimized online from the same limited pilots it uses for estimation.

%%%%%%%%%%%%%%%%%%%%%%%%%%%%%%%%%%%%%%%%%%%%%%%%%%%%%%%%%%%%%%%%%%%%%%%%%%%%%%%%%%%%%%%%%%

%\subsection{Organization and Notations}
%\label{subsec:org_notation}
\paragraph*{\textbf{Organization}}
The remainder of this paper is structured as follows. Section~\ref{Sec:SystemModel} introduces the system model and formulates pilot-limited channel estimation as an underdetermined linear inverse problem. Section~\ref{sec:GB-SMCF} then develops the GB-SMCF for array-domain prior covariance modeling. Building on these foundations, Section~\ref{Sec:proposed_framework} incorporates the GB-SMCF as a prior into the GPR-based estimator, presenting the complete CSI-reconstruction framework. Section~\ref{Sec:opt_complexity} analyzes the optimization behavior, computational complexity, and scalability of the framework to larger antenna arrays. Section~\ref{sec:simulation_results} presents numerical results validating the proposed estimator against benchmark methods. Finally, Section~\ref{sec:conclusion} concludes the paper and outlines future directions.

\paragraph*{\textbf{Notations}} Bold uppercase letters denote matrices, bold lowercase letters denote vectors, and calligraphic letters denote sets. The sets of real, complex, and integer numbers are denoted by $\R$, $\C$, and $\mathbb{Z}$, respectively. The matrix $\I_n$ denotes the $n\times n$ identity matrix, and $\mathbf{e}_i$ denotes the $i$th canonical basis vector. The operator $\vecc(\cdot)$ denotes vectorization. The superscripts $(\cdot)^{\mathsf{T}}$ and $(\cdot)^{\Hsf}$ denote transpose and Hermitian transpose, respectively. The operators $\Re\{\cdot\}$ and $\Im\{\cdot\}$ denote the real and imaginary parts, respectively, and $\mathrm{j}$ denotes the imaginary unit. The Frobenius norm is denoted by $\lVert\cdot\rVert_{\mathrm{F}}$. The trace, determinant, diagonal, and rank operators are denoted by $\tr(\cdot)$, $\det(\cdot)$, $\diag(\cdot)$, and $\rank(\cdot)$, respectively. The symbols $\otimes$ and $\circ$ denote the Kronecker and Hadamard products, respectively. The notation $\mathbf{A}\succeq\mathbf{0}$ indicates that $\mathbf{A}$ is PSD. The cardinality of a set $\mathcal{A}$ is denoted by $|\mathcal{A}|$. The distributions $\mathcal{CN}(\cdot,\cdot)$ and $\mathcal{N}(\cdot,\cdot)$ denote circularly symmetric complex Gaussian and real Gaussian distributions, respectively, and $\mathbb{E}[\cdot]$ denotes expectation. The function $k(\cdot,\cdot)$ denotes a kernel function, and $\mathbf{K}$ denotes the corresponding kernel matrix.

%%%%%%%%%%%%%%%%%%%%%%%%%%%%%%%%%%%%%%%%%%%%%%%%%%%%%%%%%%%%%%%%%%%%%%%%%%%%%%%%%%%%%%%%%%
%%%%%%%%%%%%%%%%%%%%%%%%%%%%%%%%%%%%%%%%%%%%%%%%%%%%%%%%%%%%%%%%%%%%%%%%%%%%%%%%%%%%%%%%%%
\vspace{-3mm}
\section{System Model and Problem Formulation}
\label{Sec:SystemModel}
\begin{figure}[t]
\centering
\includegraphics[width=0.8\linewidth]{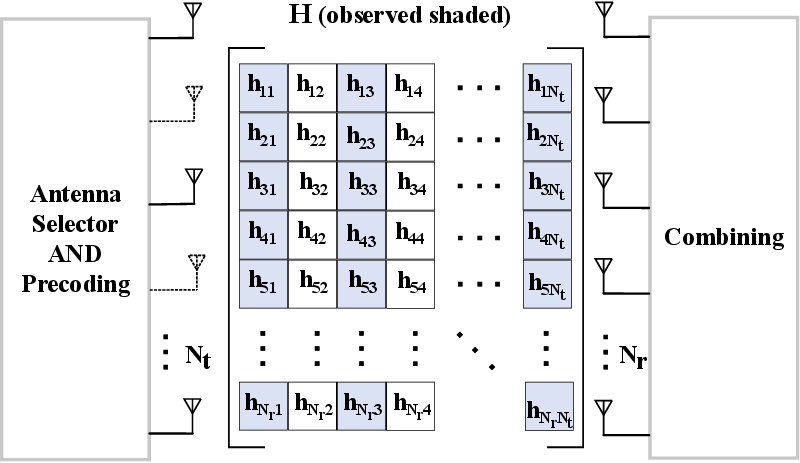}
\caption{Illustration of the proposed training architecture. A subset of transmit antennas is activated to transmit pilot symbols, while all receive antennas listen.}
\label{fig:system_model}
\end{figure}

We consider a narrowband point-to-point MIMO system, as illustrated in Fig.~\ref{fig:system_model}, where the transmitter and receiver are equipped with $N_{\textrm{t}}$ and $N_{\textrm{r}}$ antennas, respectively. To reduce pilot overhead and training energy, only $n_{\textrm{t}} \le N_{\textrm{t}}$ transmit antennas are activated during training. Specifically, for a given training phase, a known subset of transmit-antenna indices, denoted by $\Omega_{\textrm{t}}=\{a_1,\dots,a_{n_{\textrm{t}}}\}\subset\{1,\dots,N_{\textrm{t}}\}$, is selected, and only the radio-frequency chains associated with these antennas are enabled, while the remaining transmit antennas are deactivated~\cite{shah2025lowoverhead, shah2026improved}. To represent this activation pattern algebraically, we define the selection matrix $\mathbf{F}= \big[\mathbf{e}_{a_1}\ \cdots\ \mathbf{e}_{a_{n_{\textrm{t}}}}\big] \in\{0,1\}^{N_{\textrm{t}}\times n_{\textrm{t}}}$, which satisfies $\mathbf{F}^{\Hsf}\mathbf{F}=\I_{n_{\textrm{t}}}$. This matrix maps an $n_{\textrm{t}}$-dimensional pilot vector applied to the active radio-frequency chains to the full $N_{\textrm{t}}$-antenna array by placing zeros on the inactive antenna ports. Thus, for any pilot vector $\mathbf{s}\in\C^{n_{\textrm{t}}}$ applied to the active radio-frequency chains, the corresponding full-array transmit vector is $\mathbf{x}=\mathbf{F}\mathbf{s}\in\C^{N_{\textrm{t}}}$.

%%%%%%%%%%%%%%%%%%%%%%%%%%%%%%%%%%%%%%%%%%%%%%%%%%%%%%%%%%%%%%%%%%%%%%%%%%%%%%%%%%%%%%%%%%
\vspace{-2mm}
\subsection{Training Signal Model}
Let $\mathbf{S}\in\C^{n_{\textrm{t}}\times{T}}$ denote the pilot matrix transmitted over $T$ channel uses within a coherence block, with the $t$-th column $\mathbf{s}_{t}$ sent at time $t$. Each active antenna transmits with fixed per-symbol pilot power $P_{\textrm{A}}>0$, so the transmitted pilot matrix is $\mathbf{X}_{\textrm{TX}}=\sqrt{P_{\textrm{A}}}\,\mathbf{F}\mathbf{S}\in\C^{N_{\textrm{t}}\times T}$. We adopt constant-modulus row-orthogonal pilots satisfying $\mathbf{S}\mathbf{S}^{\Hsf}=T\I_{n_{\textrm{t}}}$ and $|[\mathbf{S}]_{\ell,t}|=1$ for all $\ell\in\{1,\ldots,n_{\textrm{t}}\}$ and $t\in\{1,\ldots,T\}$, so the total training energy over one coherence block is $n_{\textrm{t}}TP_{\textrm{A}}$.

Let $\mathbf{H}\in\C^{N_{\textrm{r}}\times N_{\textrm{t}}}$ denote the unknown narrowband channel, and let $\mathbf{N}\in\C^{N_{\textrm{r}}\times T}$ denote the additive noise matrix whose entries are independent and identically distributed as $\CN(0,\sigma_n^2)$. The received pilot matrix is $\mathbf{Y}_{\textrm{RX}}=\mathbf{H}\mathbf{X}_{\textrm{TX}}+\mathbf{N}=\sqrt{P_{\textrm{A}}}\,\mathbf{H}\mathbf{F}\mathbf{S}+\mathbf{N}\in\C^{N_{\textrm{r}}\times T}$, which depends on $\mathbf{H}$ only through the selected columns $\mathbf{H}\mathbf{F}$. Correlating $\mathbf{Y}_{\textrm{RX}}$ with $\mathbf{S}$ and normalizing by $T$ gives $\widetilde{\mathbf{Y}}_{\textrm{RX}}= \frac{1}{T}\mathbf{Y}_{\textrm{RX}}\mathbf{S}^{\Hsf}=\sqrt{P_{\textrm{A}}}\,\mathbf{H}\mathbf{F}+\widetilde{\mathbf{N}}$, where $\widetilde{\mathbf{N}}=\frac{1}{T}\mathbf{N}\mathbf{S}^{\Hsf}\in\C^{N_{\textrm{r}}\times n_{\textrm{t}}}$. Since $\mathbf{S}\mathbf{S}^{\Hsf}=T\I_{n_{\textrm{t}}}$, the active-antenna observations are separated without inter-pilot interference, and $\vecc(\widetilde{\mathbf{N}})\sim \CN\!\Big(\mathbf{0},\frac{\sigma_n^2}{T}\I_{N_{\textrm{r}}n_{\textrm{t}}}\Big)$. Defining $\mathbf{Z}= \widetilde{\mathbf{Y}}_{\textrm{RX}}/\sqrt{P_{\textrm{A}}}$, we obtain
\begin{equation}
\label{eq:Zdef}
\mathbf{Z}
=
\mathbf{H}\mathbf{F}+\mathbf{W}\in\C^{N_{\textrm{r}}\times n_{\textrm{t}}},
\;
\vecc(\mathbf{W})
\sim
\CN\Big(\mathbf{0},\frac{\sigma_n^2}{TP_{\textrm{A}}}\I_{N_{\textrm{r}}n_{\textrm{t}}}\Big).
\end{equation}
%%%%%%%%%%%%%%%%%%%%%%%%%%%%%%%%%%%%%%%%%%%%%%%%%%%%%%%%%%%%%%%%%%%%%%%%%%%%%%%%%%%%%%%%%%

\vspace{-2mm}
\subsection{Problem Formulation}
The receiver therefore observes $\mathbf{Z}\in\C^{N_{\textrm{r}}\times n_{\textrm{t}}}$, which is a noisy version of the channel columns selected by $\Omega_{\textrm{t}}$. To express~\eqref{eq:Zdef} in vector form, we define $\mathbf{z}=\vecc(\mathbf{Z})$, $\mathbf{h}=\vecc(\mathbf{H})$, and $\mathbf{w}=\vecc(\mathbf{W})$, and let
$\boldsymbol{\Phi}= \mathbf{F}^{\mathsf{T}}\otimes \I_{N_{\textrm{r}}}\in\{0,1\}^{N_{\textrm{r}}n_{\textrm{t}}\times N_{\textrm{r}}N_{\textrm{t}}}$
be the corresponding channel-selection matrix. Using
$\vecc(\mathbf{H}\mathbf{F})=(\mathbf{F}^{\mathsf{T}}\otimes \I_{N_{\textrm{r}}})\vecc(\mathbf{H})$, the observation model becomes
\begin{equation}
\label{eq:vecModel}
\mathbf{z}
=
\boldsymbol{\Phi}\mathbf{h}+\mathbf{w}\in\C^{N_{\textrm{r}}n_{\textrm{t}}}.
\end{equation}
Here, $\boldsymbol{\Phi}$ extracts exactly those entries of $\mathbf{h}$ whose transmit-antenna indices belong to $\Omega_{\textrm{t}}$. Since the number of observations is $N_{\textrm{r}}n_{\textrm{t}}$ whereas the number of unknowns is $N_{\textrm{r}}N_{\textrm{t}}$, the problem is highly underdetermined when $n_{\textrm{t}}\ll N_{\textrm{t}}$.

\paragraph*{\textbf{Objective}}
Given the observation vector $\mathbf{z}$ in~\eqref{eq:vecModel}, the objective is to reconstruct the channel matrix $\mathbf{H}\in\C^{N_{\textrm{r}}\times N_{\textrm{t}}}$. Since $\mathbf{z}$ contains noisy measurements only from the transmit-antenna columns indexed by $\Omega_{\textrm{t}}$, the unobserved channel coefficients cannot be recovered independently. Their reconstruction requires a statistical model that couples the observed and unobserved entries across the joint transmit--receive antenna-index lattice. In the following sections, this coupling is modeled through a proper-complex GP prior whose covariance is specified by the proposed GB-SMCF.

%%%%%%%%%%%%%%%%%%%%%%%%%%%%%%%%%%%%%%%%%%%%%%%%%%%%%%%%%%%%%%%%%%%%%%%%%%%%%%%%%%%%%%%%%%
%%%%%%%%%%%%%%%%%%%%%%%%%%%%%%%%%%%%%%%%%%%%%%%%%%%%%%%%%%%%%%%%%%%%%%%%%%%%%%%%%%%%%%%%%%
\vspace{-2mm}
\section{The Proposed Geometry-Aware Kernel}
\label{sec:GB-SMCF}
Accurate CSI acquisition under limited pilot and energy budgets requires a prior that captures the spatial correlation structure of the wireless channel~\cite{shah2026improved,shah2025lowoverhead}. The classical Kronecker model~\cite{kronecker2002} is widely used because of its simplicity, but it assumes separability of the receive and transmit correlation structures, which may fail in realistic propagation environments, particularly under near-field effects, dual-polarized arrays, or rich scattering~\cite{Jieao2025IEEE_TIT}. Indeed, the Kronecker model has been shown to underestimate channel capacity and misrepresent eigenvalue distributions in measured indoor scenarios~\cite{Ozcelik2003}. Fixed-kernel approaches, such as SqExp or Mat\'ern kernels, can model isotropic or anisotropic decay, but they cannot represent the oscillatory structure induced by constructive and destructive multipath interference~\cite{shah2025lowoverhead}. To address these limitations, this section introduces the proposed GB-SMCF kernel for the GPR-based CSI reconstruction framework in Section~\ref{Sec:proposed_framework}.
%%%%%%%%%%%%%%%%%%%%%%%%%%%%%%%%%%%%%%%%%%%%%%%%%%%%%%%%%%%%%%%%%%%%%%%%%%%%%%%%%%%%%%%%%%
\vspace{-2mm}
\subsection{Proper Complex GP Modeling of the Channel}
To capture statistical coupling among the channel coefficients, we model the channel as a zero-mean proper complex Gaussian random field over the joint transmit--receive antenna-index grid. Specifically, define the antenna-index grid $\mathcal{G} = \big\{(i,j): i\in\{1,\dots,N_{\textrm{r}}\},\ j\in\{1,\dots,N_{\textrm{t}}\}\big\}$, and let $h:\mathcal{G}\to\C$ denote the corresponding channel field, with $h(i,j)=H_{i,j}$. Thus, within one coherence block, the channel matrix $\mathbf{H}\in\C^{N_{\textrm{r}}\times N_{\textrm{t}}}$ collects the field values over $\mathcal{G}$. We place a complex GP prior on $h$ as $h \sim \mathcal{GP}_{\C}\big(0,k_{\textrm{GB}}(\cdot,\cdot)\big)$, where $k_{\textrm{GB}}\big((i,j),(i',j')\big)$ is the GB-SMCF covariance function describing the statistical coupling between $h(i,j)$ and $h(i',j')$. Hence, for any finite collection of channel coefficients, the prior induces a complex Gaussian vector $\mathbf{h}\sim\mathcal{CN}(\mathbf{0},\mathbf{K}_{\textrm{GB}})$, whose entries are given by $k_{\textrm{GB}}(\cdot,\cdot)$. The GB-SMCF covariance function is introduced next in Subsection~\ref{SubSec:GB-SMCF}. Under this prior, conditioning on the observed coefficients yields a Gaussian posterior for the unobserved ones, making Bayesian CSI reconstruction analytically tractable~\cite{rasmussen2006gaussian, Jieao2025IEEE_TIT}.

For subsequent inference, we adopt an equivalent real-coordinate representation of this finite-dimensional prior. Specifically, if $\mathbf{h}\sim\mathcal{CN}(\mathbf{0},\mathbf{K}_{\textrm{GB}})$ is any finite channel vector drawn from the GP prior and $\mathbf{f}=[\Re\{\mathbf{h}\}^{\mathsf{T}}\ \Im\{\mathbf{h}\}^{\mathsf{T}}]^{\mathsf{T}}$, then $\mathbf{f}$ is real Gaussian with covariance
\begin{equation}
\label{eq:T-operator}
\mathcal{T}(\mathbf{K}_{\textrm{GB}})
=
\frac{1}{2}
\begin{bmatrix}
\Re\{\mathbf{K}_{\textrm{GB}}\} & -\Im\{\mathbf{K}_{\textrm{GB}}\}\\
\Im\{\mathbf{K}_{\textrm{GB}}\} & \Re\{\mathbf{K}_{\textrm{GB}}\}
\end{bmatrix}.
\end{equation}
For rectangular cross-covariances, the same transformation applies: if $\mathbf{K}_{\mathcal{X}\mathcal{X}_\star}$ is the complex covariance between finite index sets $\mathcal{X}$ and $\mathcal{X}_\star$, then $\mathcal{T}(\mathbf{K}_{\mathcal{X}\mathcal{X}_\star})$ is the corresponding real-coordinate cross-covariance between their stacked real--imaginary representations. Thus, $\mathcal{T}(\cdot)$ provides an exact real-coordinate representation of the proper complex covariance. This form is used only for the posterior computations in Section~\ref{Sec:proposed_framework}; the channel prior itself remains a proper complex GP defined directly by $k_{\textrm{GB}}(\cdot,\cdot)$.

%%%%%%%%%%%%%%%%%%%%%%%%%%%%%%%%%%%%%%%%%%%%%%%%%%%%%%%%%%%%%%%%%%%%%%%%%%%%%%%%%%%%%%%%%%
\vspace{-3mm}
\subsection{Construction of the GB-SMCF}
\label{SubSec:GB-SMCF}
We construct the GB-SMCF by introducing the URA lattice geometry and index differences, followed by the definition of a separable complex spectral-mixture covariance on the joint transmit--receive index lattice~\cite{wilson2013gaussian_SM_kernel}.

\subsubsection{Array Lattice and Index Differences}
Consider receive and transmit URAs with $N_{\textrm{r}}=N_y^{(\textrm{r})}N_z^{(\textrm{r})}$ and $N_{\textrm{t}}=N_y^{(\textrm{t})}N_z^{(\textrm{t})}$ elements, respectively. Through a fixed bijection between array-element indices and lattice coordinates, each receive index $i$ and transmit index $j$ is associated with centered coordinates
\begin{equation}
\label{eq:lattice_coords}
\mathbf{r}_i=(y_i^{(\textrm{r})},z_i^{(\textrm{r})})\in\mathbb{Z}^2, \;
\mathbf{t}_j=(y_j^{(\textrm{t})},z_j^{(\textrm{t})})\in\mathbb{Z}^2.
\end{equation}
For antenna pairs $(i,i')$ and $(j,j')$, define the receive- and transmit-side lattice differences as $\Delta\mathbf{r}= \mathbf{r}_i-\mathbf{r}_{i'}=(\Delta r_y,\Delta r_z)$ and $\Delta\mathbf{t}= \mathbf{t}_j-\mathbf{t}_{j'}=(\Delta t_y,\Delta t_z)$. Under lattice stationarity, the covariance depends on the antenna indices only through the relative displacements $(\Delta\mathbf{r},\Delta\mathbf{t})$, rather than on their absolute locations. This makes the kernel consistent with the shift-invariant second-order structure induced by the URA geometry.

\subsubsection{Separable Complex Spectral-Mixture Kernel on the Array Index Domain}
Motivated by the spectral characterization of stationary kernels~\cite{wilson2013gaussian_SM_kernel}, we model the receive- and transmit-side correlations by complex spectral-mixture covariances on their respective URA lattices. For the receive side, define
\begin{equation}
\label{eq:K-rx}
\begin{aligned}
k_{\textrm{r}}(i,i')
&=
\sum_{q=1}^{Q_{\textrm{r}}}
\pi_q^{(\textrm{r})}
\exp\!\Big(-2\pi^2\big[v_{q,y}^{(\textrm{r})}(\Delta r_y)^2+v_{q,z}^{(\textrm{r})}(\Delta r_z)^2\big]\Big)\\
&\qquad\times
\exp\!\Big(\mathrm{j}\,2\pi\big[\mu_{q,y}^{(\textrm{r})}\Delta r_y+\mu_{q,z}^{(\textrm{r})}\Delta r_z\big]\Big),
\end{aligned}
\end{equation}
and for the transmit side,
\begin{equation}
\label{eq:K-tx}
\begin{aligned}
k_{\textrm{t}}(j,j')
&=
\sum_{p=1}^{Q_{\textrm{t}}}
\pi_p^{(\textrm{t})}
\exp\!\Big(-2\pi^2\big[v_{p,y}^{(\textrm{t})}(\Delta t_y)^2+v_{p,z}^{(\textrm{t})}(\Delta t_z)^2\big]\Big)\\
&\qquad\times
\exp\!\Big(\mathrm{j}\,2\pi\big[\mu_{p,y}^{(\textrm{t})}\Delta t_y+\mu_{p,z}^{(\textrm{t})}\Delta t_z\big]\Big).
\end{aligned}
\end{equation}
The proposed GB-SMCF on the joint antenna-index domain is then defined by the separable product
\begin{equation}
\label{eq:GB-SM-function}
k_{\textrm{GB}}\big((i,j),(i',j')\big)
=
\alpha\,k_{\textrm{r}}(i,i')\,k_{\textrm{t}}(j,j'),
\end{equation}
where $\alpha>0$ is a global variance parameter. The receive- and transmit-side mixture weights satisfy $\pi_q^{(\textrm{r})}\ge 0$, $\sum_{q=1}^{Q_{\textrm{r}}}\pi_q^{(\textrm{r})}=1$, $\pi_p^{(\textrm{t})}\ge 0$, and $\sum_{p=1}^{Q_{\textrm{t}}}\pi_p^{(\textrm{t})}=1$, while the spectral-variance parameters satisfy $v_{q,\xi}^{(\textrm{r})}>0$ and $v_{p,\xi}^{(\textrm{t})}>0$ for $\xi\in\{y,z\}$. With this normalization, $k_{\textrm{r}}(i,i)=k_{\textrm{t}}(j,j)=1$ for all $i,j$, and hence $k_{\textrm{GB}}\big((i,j),(i,j)\big)=\alpha$. Therefore, $\alpha$ alone sets the prior marginal variance, while the mixture weights $\pi_q^{(\cdot)}$ and $\pi_p^{(\cdot)}$ control only the relative spectral allocation of the components. The separable form preserves transmit--receive tractability while allowing each side to capture multiple oscillatory modes together with smooth spatial decay.

\begin{lemma}[Hermitian PSD and Lattice Stationarity]
\label{lem:PSD}
For normalized nonnegative weights $\{\pi_q^{(\textrm{r})}\}$ and $\{\pi_p^{(\textrm{t})}\}$, spectral-variance parameters $v_{q,\xi}^{(\textrm{r})},v_{p,\xi}^{(\textrm{t})}>0$, and $\alpha>0$, the kernel $k_{\textrm{GB}}(\cdot,\cdot)$ in~\eqref{eq:K-rx}--\eqref{eq:GB-SM-function} is Hermitian PSD on $\mathcal{G}\times\mathcal{G}$ and depends only on the lattice differences $(\Delta\mathbf{r},\Delta\mathbf{t})$. Moreover, for any finite ordered set $\mathcal{A}\subset\mathcal{G}$, the real-augmented covariance matrix $\mathcal{T}(\mathbf{K}_{\textrm{GB}}(\mathcal{A},\mathcal{A}))$ is PSD, and adding isotropic observation-noise covariance preserves positive semidefiniteness.
\end{lemma}
\begin{IEEEproof}
See Appendix~\ref{app:psd-proof}.
\end{IEEEproof}

\subsubsection{Interpretation of the GB-SMCF Hyperparameters}
Each mixture component in the receive- and transmit-side kernels represents a spatial spectral component on the corresponding URA lattice. On each side, the normalized weights $\pi_q^{(\textrm{r})}$ and $\pi_p^{(\textrm{t})}$ allocate relative spectral mass across components; the normalized spatial frequencies $\mu_{q,y}^{(\textrm{r})},\mu_{q,z}^{(\textrm{r})}$ and $\mu_{p,y}^{(\textrm{t})},\mu_{p,z}^{(\textrm{t})}$ determine the oscillatory behavior along the two lattice axes; and the spectral-variance parameters $v_{q,y}^{(\textrm{r})},v_{q,z}^{(\textrm{r})}$ and $v_{p,y}^{(\textrm{t})},v_{p,z}^{(\textrm{t})}$ govern the correlation-decay rate with lattice separation, equivalently setting the correlation lengths and effective angular spreads along those axes. The scalar $\alpha>0$ scales the overall covariance level and, by the diagonal normalization of the side kernels, equals the prior marginal variance of each channel coefficient.

%%%%%%%%%%%%%%%%%%%%%%%%%%%%%%%%%%%%%%%%%%%%%%%%%%%%%%%%%%%%%%%%%%%%%%%%%%%%%%%%%%%%%%%%%%
\vspace{-2mm}
\subsection{Properties of the GB-SMCF}
\label{subsec:Prop_GB-SMCF}
The GB-SMCF encodes the URA lattice structure as a proper complex covariance kernel on the discrete antenna-index domain.

By construction, $k_{\textrm{GB}}(\cdot,\cdot)$ depends only on the lattice differences $(\Delta\mathbf{r},\Delta\mathbf{t})$, and therefore defines a stationary prior on $\mathcal{G}$. The factorization $k_{\textrm{GB}}=\alpha\,k_{\textrm{r}}k_{\textrm{t}}$ preserves transmit--receive separability, yielding a Kronecker-like structure while retaining a rich parametric form on each side. Since $k_{\textrm{r}}$ and $k_{\textrm{t}}$ are two-dimensional complex spectral-mixture kernels, the prior can represent multiple angular components through distinct spectral locations and widths~\cite{wilson2013gaussian_SM_kernel}. Unlike purely distance-based kernels such as the SqExp or Mat\'ern kernels, this form captures both oscillatory behavior and spatial decay~\cite{shah2026improved}. The parameters $v_{\cdot,y}^{(s)}$ and $v_{\cdot,z}^{(s)}$ further allow different correlation lengths along the two URA axes, thereby capturing azimuth--elevation anisotropy on both the transmit and receive sides.

The model is specified directly in the complex domain, so the coupling between the real and imaginary parts is induced by the Hermitian covariance kernel rather than via an auxiliary output-correlation construction. The real-coordinate representation obtained through $\mathcal{T}(\cdot)$ is exact; hence, inference with the stacked real--imaginary form is algebraically equivalent to conditioning under the proper complex prior. Because the kernel is defined on the full lattice $\mathcal{G}$, posterior inference extends naturally from the observed subset to unobserved in-grid antenna pairs. The hyperparameters retain clear physical interpretations; weights $\pi$ for spectral mass, frequencies $\mu$ for spatial carriers, variances $v$ for correlation scales, and $\alpha$ for overall variance. All hyperparameters are learned directly from sparse pilot observations by marginal-likelihood maximization, without offline training~\cite{WeiBer2026,Wei2021deep,Fesl2024enhancing}. Finally, Lemma~\ref{lem:PSD} guarantees that the kernel is Hermitian PSD and that the corresponding real-augmented covariance remains PSD after adding isotropic observation noise. Since the hyperparameters are re-estimated in each coherence block, the model adapts to blockwise changes in propagation statistics while remaining stationary within each block. Explicit spatial non-stationarity across the array aperture is beyond the present formulation and may be incorporated through non-stationary extensions of $k_{\textrm{GB}}(\cdot,\cdot)$~\cite{Zhang2026}.

%%%%%%%%%%%%%%%%%%%%%%%%%%%%%%%%%%%%%%%%%%%%%%%%%%%%%%%%%%%%%%%%%%%%%%%%%%%%%%%%%%%%%%%%%%
%%%%%%%%%%%%%%%%%%%%%%%%%%%%%%%%%%%%%%%%%%%%%%%%%%%%%%%%%%%%%%%%%%%%%%%%%%%%%%%%%%%%%%%%%%
\vspace{-2mm}
\section{Proposed Framework}
\label{Sec:proposed_framework}
We now combine the observation model with the GB-SMCF prior to reconstruct the full CSI from sparse noisy pilot observations. The proposed framework proceeds in three steps: first, the received pilot observations are organized into observed and unobserved index sets over the antenna-index lattice; second, the GB-SMCF hyperparameters are learned by marginal-likelihood maximization; third, the learned prior is conditioned on the observations to obtain the posterior distribution and the resulting CSI estimate. Fig.~\ref{fig:gpr_csi_pipeline} illustrates the overall pipeline.

\begin{figure*}[t]
\centering
%\resizebox{\columnwidth}{!}{%
\begin{tikzpicture}[
  >=Latex,
  node distance=6mm,
  box/.style = {
    draw, rounded corners, thick, align=center, inner sep=1mm,
    minimum width=3.5cm, font=\footnotesize
  },
  arrow/.style = {
    -{Latex[length=3.5mm,width=2mm]}, thick
  }
]

\node[box, fill=blue!10] (tx) {%
\textbf{Transmitter}\\
$\mathbf{F}=[\mathbf{e}_{a_1}\ \cdots\ \mathbf{e}_{a_{n_{\textrm{t}}}}]$\\
$\mathbf{S}\in\C^{n_{\textrm{t}}\times T},\ \mathbf{S}\mathbf{S}^{\Hsf}=T\I_{n_{\textrm{t}}}$\\
$\mathbf{X}_{\textrm{TX}}=\sqrt{P_{\textrm{A}}}\mathbf{F}\mathbf{S}$
};

\node[box, fill=gray!8, right=of tx] (ch) {%
\textbf{Channel}\\
$\mathbf{H}\in\C^{N_{\textrm{r}}\times N_{\textrm{t}}}$
};

\node[box, fill=green!10, right=of ch] (rx) {%
\textbf{Receiver}\\
$\mathbf{Y}_{\textrm{RX}}=\mathbf{H}\mathbf{X}_{\textrm{TX}}+\mathbf{N}$\\
$\widetilde{\mathbf{Y}}_{\textrm{RX}}=\frac{1}{T}\mathbf{Y}_{\textrm{RX}}\mathbf{S}^{\Hsf}$,\,
$\mathbf{Z}=\widetilde{\mathbf{Y}}_{\textrm{RX}}/\sqrt{P_{\textrm{A}}}$\\
$\mathbf{z}=\vecc(\mathbf{Z})$,\;
$\mathbf{z}_{\textrm{aug}}=\big[\Re\{\mathbf{z}\}^{\mathsf{T}}\ \Im\{\mathbf{z}\}^{\mathsf{T}}\big]^{\mathsf{T}}$
};

\draw[arrow] (tx) -- (ch);
\draw[arrow] (ch) -- (rx);

\node[box, fill=orange!10, below= of rx, yshift=-3mm,] (prior) {%
\textbf{GB-SMCF Prior}\\
$k_{\textrm{GB}}((i,j),(i',j'))=\alpha\,k_{\textrm{r}}(i,i')\,k_{\textrm{t}}(j,j')$\\
$\mathbf{K}_{\textrm{GB}}(\mathcal{X},\mathcal{X})$ from $k_{\textrm{GB}}$\\
$\widetilde{\mathbf{K}}_{\mathcal{X}\mathcal{X}}=\mathcal{T}\!\big(\mathbf{K}_{\textrm{GB}}(\mathcal{X},\mathcal{X})\big)$
};

\node[box, fill=yellow!10, left=of prior] (post) {%
\textbf{Posterior Update}\\
$\boldsymbol{m}_{\mathcal{X}}=\widetilde{\mathbf{K}}_{\mathcal{X}\mathcal{X}}\mathbf{C}_\theta^{-1}\mathbf{z}_{\textrm{aug}}$\\
$\boldsymbol{m}_{\mathcal{X}_\star}=\widetilde{\mathbf{K}}_{\mathcal{X}_\star\mathcal{X}}\mathbf{C}_\theta^{-1}\mathbf{z}_{\textrm{aug}}$\\
$\boldsymbol{\Sigma}_\star=\widetilde{\mathbf{K}}_{\mathcal{X}_\star\mathcal{X}_\star}-\widetilde{\mathbf{K}}_{\mathcal{X}_\star\mathcal{X}}\mathbf{C}_\theta^{-1}\widetilde{\mathbf{K}}_{\mathcal{X}_\star\mathcal{X}}^{\mathsf{T}}$
};

\node[box, fill=red!10, left=of post] (est) {%
\textbf{Estimated CSI}\\
$\widehat{\mathbf{h}}_{\mathcal{X}}=\boldsymbol{m}_{\mathcal{X}}^{\Re}+\mathrm{j}\,\boldsymbol{m}_{\mathcal{X}}^{\Im}$\\
$\widehat{\mathbf{h}}_{\mathcal{X}_\star}=\boldsymbol{m}_{\mathcal{X}_\star}^{\Re}+\mathrm{j}\,\boldsymbol{m}_{\mathcal{X}_\star}^{\Im}$\\
Assemble $\widehat{\mathbf{H}}_{\textrm{GPR}}\in\C^{N_{\textrm{r}}\times N_{\textrm{t}}}$
};

\draw[arrow] (rx) -- (prior);
\draw[arrow] (prior) -- (post);
\draw[arrow] (post) -- (est);

\draw[arrow, dashed]
  (prior.north) to[bend right=20]
  node[midway, right, yshift=2mm, xshift=-30mm, font=\scriptsize]
  {Learned hyperparameters}
  (post.north);

\end{tikzpicture}%
%}
\caption{The proposed GB-SMCF-based GPR framework for pilot-limited channel estimation under a proper complex GP prior.}
\label{fig:gpr_csi_pipeline}
\end{figure*}

%%%%%%%%%%%%%%%%%%%%%%%%%%%%%%%%%%%%%%%%%%%%%%%%%%%%%%%%%%%%%%%%%%%%%%%%%%%%%%%%%%%%%%%%%%
\vspace{-2mm}
\subsection{Observation Model}
\label{SubSec:Observation_DataSet}
We derive the observed and unobserved index sets for GPR from the received noisy pilot observation model in~\eqref{eq:Zdef}. Activating the transmit-antenna subset indexed by $\Omega_{\textrm{t}}=\{a_1,\dots,a_{n_{\textrm{t}}}\}\subset\{1,\dots,N_{\textrm{t}}\}$ yields noisy observations of the channel coefficients on the index set
\begin{equation}
\label{eq:X_train_pf}
\mathcal{X}=\{x_p\}_{p=1}^P
=
\{(i,a_\ell):\, i=1,\dots,N_{\textrm{r}},\ \ell=1,\dots,n_{\textrm{t}}\}\subset\mathcal{G},
\end{equation}
with cardinality $|\mathcal{X}|=P=N_{\textrm{r}}n_{\textrm{t}}$. The complementary set of unobserved channel indices is
\begin{equation}
\label{eq:Xstar_pf}
\mathcal{X}_\star=\mathcal{G}\setminus\mathcal{X},
\qquad
|\mathcal{X}_\star|=P_\star.
\end{equation}
For each observed index $x_p=(i,a_\ell)\in\mathcal{X}$, the corresponding entry of $\mathbf{Z}$ in~\eqref{eq:Zdef} satisfies
\begin{equation}
\label{eq:obs_scalar_pf}
z_p=z(i,a_\ell)=h(i,a_\ell)+w_p,
\qquad
w_p\sim\mathcal{CN}(0,\sigma_{\textrm{obs}}^2),
\end{equation}
with $w_p$ independent across $p$, where $\sigma_{\textrm{obs}}^2=\sigma_n^2/(TP_{\textrm{A}})$. Thus, $\mathbf{z}=\vecc(\mathbf{Z})\in\C^{P}$ in~\eqref{eq:vecModel} is precisely the noisy observation vector on $\mathcal{X}$.

Its real-coordinate form is
\begin{equation}
\label{eq:z_icm}
\mathbf{z}_{\textrm{aug}}
=
[\Re\{\mathbf{z}\}^{\mathsf{T}}\ \Im\{\mathbf{z}\}^{\mathsf{T}}]^{\mathsf{T}}
=
\mathbf{f}_{\mathcal{X}}+\mathbf{n},
\;
\mathbf{n}\sim\mathcal{N}\!\Big(\mathbf{0},\frac{\sigma_{\textrm{obs}}^2}{2}\I_{2P}\Big).
\end{equation}
Here, $\mathbf{h}_{\mathcal{X}}=[h(x_1),\dots,h(x_P)]^{\mathsf{T}}\in\C^{P}$ and $\mathbf{f}_{\mathcal{X}}=[\Re\{\mathbf{h}_{\mathcal{X}}\}^{\mathsf{T}}\ \Im\{\mathbf{h}_{\mathcal{X}}\}^{\mathsf{T}}]^{\mathsf{T}}\in\R^{2P}$ are the complex and real-augmented observed channel vectors, respectively. Similarly, for the unobserved set $\mathcal{X}_\star=\{x_{\star,q}\}_{q=1}^{P_\star}$, define $\mathbf{h}_{\mathcal{X}_\star}=[h(x_{\star,1}),\dots,h(x_{\star,P_\star})]^{\mathsf{T}}\in\C^{P_\star}$ and $\mathbf{f}_{\mathcal{X}_\star}=[\Re\{\mathbf{h}_{\mathcal{X}_\star}\}^{\mathsf{T}}\ \Im\{\mathbf{h}_{\mathcal{X}_\star}\}^{\mathsf{T}}]^{\mathsf{T}}\in\R^{2P_\star}$. Hence, $\mathbf{f}_{\mathcal{X}}$ and $\mathbf{f}_{\mathcal{X}_\star}$ collect the real and imaginary parts of the observed and unobserved channel coefficients, respectively, providing the training data on $\mathcal{X}$ and the prediction targets on $\mathcal{X}_\star$ for the subsequent GPR framework.

%%%%%%%%%%%%%%%%%%%%%%%%%%%%%%%%%%%%%%%%%%%%%%%%%%%%%%%%%%%%%%%%%%%%%%%%%%%%%%%%%%%%%%%%%%
\vspace{-3mm}
\subsection{GB-SMCF Hyperparameter Learning}
\label{subsec:kernel-learning}
Using the training data $(\mathcal{X},\mathbf{z}_{\textrm{aug}})$, we learn the GB-SMCF hyperparameters by maximizing the log marginal likelihood induced by the proper complex GP prior. We collect the model hyperparameters from~\eqref{eq:K-rx}--\eqref{eq:GB-SM-function} as
\begin{equation}
\begin{aligned}
\label{eq:kernel_hyperparameters}
\vartheta\equiv\big\{&
\alpha;\,
\{\pi_q^{(\textrm{r})},\mu_{q,y}^{(\textrm{r})},\mu_{q,z}^{(\textrm{r})},v_{q,y}^{(\textrm{r})},v_{q,z}^{(\textrm{r})}\}_{q=1}^{Q_{\textrm{r}}};\\
&
\{\pi_p^{(\textrm{t})},\mu_{p,y}^{(\textrm{t})},\mu_{p,z}^{(\textrm{t})},v_{p,y}^{(\textrm{t})},v_{p,z}^{(\textrm{t})}\}_{p=1}^{Q_{\textrm{t}}}
\big\}.
\end{aligned}
\end{equation}
Let $\theta$ denote the optimization variables, with $\vartheta=\vartheta(\theta)$ denoting the corresponding kernel parameters in~\eqref{eq:GB-SM-function}. To optimize over $\theta$ while enforcing the native parameter domains on each side $s\in\{\textrm{r},\textrm{t}\}$, we use the reparameterization
\begin{equation}
\label{eq:reparam}
\begin{aligned}
\alpha &= \exp(\theta_\alpha), \\[1mm]
\pi_m^{(s)}
&=
\frac{\exp(\theta_{\pi,m}^{(s)})}
{1+\sum_{\ell=1}^{Q_s-1}\exp(\theta_{\pi,\ell}^{(s)})},
\qquad m=1,\dots,Q_s-1,\\[1mm]
\pi_{Q_s}^{(s)}
&=
\frac{1}
{1+\sum_{\ell=1}^{Q_s-1}\exp(\theta_{\pi,\ell}^{(s)})},
\; s\in\{\textrm{r},\textrm{t}\},\\[1mm]
\mu_{m,\xi}^{(s)} &= \tfrac{1}{2}\tanh(\theta_{\mu,m,\xi}^{(s)}),
\;
v_{m,\xi}^{(s)} = \exp(\theta_{v,m,\xi}^{(s)}),
\; \xi\in\{y,z\}.
\end{aligned}
\end{equation}
With $\widetilde{\mathbf{K}}_{\mathcal{X}\mathcal{X}}
=\mathcal{T}\!\big(\mathbf{K}_{\textrm{GB}}(\mathcal{X},\mathcal{X})\big)$
and $\boldsymbol{\Sigma}_{\textrm{n}}
=(\sigma_{\textrm{obs}}^2/2)\I_{2P}$, the real-augmented observation vector in~\eqref{eq:z_icm} is Gaussian under the zero-mean proper complex GP prior, with covariance
$\mathbf{C}_\theta
=\widetilde{\mathbf{K}}_{\mathcal{X}\mathcal{X}}
+\boldsymbol{\Sigma}_{\textrm{n}}$. The log marginal likelihood is
\begin{equation}
\label{eq:LML}
\mathcal{L}(\theta)
=
-\frac{1}{2}\mathbf{z}_{\textrm{aug}}^{\mathsf{T}}\mathbf{C}_\theta^{-1}\mathbf{z}_{\textrm{aug}}
-\frac{1}{2}\log\det\mathbf{C}_\theta
-P\log(2\pi).
\end{equation}
Accordingly, hyperparameter learning is posed as
\begin{equation}
\label{eq:opt-problem}
\begin{aligned}
&\underset{\theta}{\text{maximize}} \quad && \mathcal{L}(\theta)\\
&\text{subject to} \quad && \theta\in\Theta \ \text{via~\eqref{eq:reparam}},
\end{aligned}
\end{equation}
where $\Theta$ denotes the box-constrained domain for $\theta$ corresponding, through~\eqref{eq:reparam}, to the hyperparameter ranges specified in Table~\ref{tab:sim_params}. Let $\mathbf{q}_\theta=\mathbf{C}_\theta^{-1}\mathbf{z}_{\textrm{aug}}$. Then, for the $m$th component $\theta_m$ of $\theta$,
\begin{equation}
\label{eq:GP-grad-master}
\frac{\partial\mathcal{L}}{\partial\theta_m}
=
\frac{1}{2}\tr\!\Big[
\big(\mathbf{q}_\theta\mathbf{q}_\theta^{\mathsf{T}}-\mathbf{C}_\theta^{-1}\big)
\frac{\partial \mathbf{C}_\theta}{\partial\theta_m}
\Big].
\end{equation}
The required derivatives inherit the separable structure of the GB-SMCF kernel. Since $\mathcal{T}(\cdot)$ is linear, for each component $\theta_m$ of $\theta$,
$
\frac{\partial\mathbf{C}_\theta}{\partial\theta_m}
=
\mathcal{T}\!\left(
\frac{\partial\mathbf{K}_{\textrm{GB}}(\mathcal{X},\mathcal{X})}{\partial\theta_m}
\right).
$
In particular,
\begin{equation}
\resizebox{\columnwidth}{!}{$
\label{eq:dKGB-main}
\frac{\partial \mathbf{K}_{\textrm{GB}}}{\partial \theta_\alpha}
=
\mathbf{K}_{\textrm{GB}},
\;
\frac{\partial \mathbf{K}_{\textrm{GB}}}{\partial \theta_{\textrm{r}}}
=
\alpha\Big(\frac{\partial\mathbf{K}_{\textrm{r}}}{\partial \theta_{\textrm{r}}}\circ \mathbf{K}_{\textrm{t}}\Big),
\;
\frac{\partial \mathbf{K}_{\textrm{GB}}}{\partial \theta_{\textrm{t}}}
=
\alpha\Big(\mathbf{K}_{\textrm{r}}\circ\frac{\partial \mathbf{K}_{\textrm{t}}}{\partial \theta_{\textrm{t}}}\Big),
$}
\end{equation}
where $\theta_{\textrm{r}}$ and $\theta_{\textrm{t}}$ denote arbitrary receive-side and transmit-side optimization variables, respectively. The closed-form side-kernel derivatives are given in Appendix~\ref{app:kernel-grads}. Algorithm~\ref{alg:cov_design} summarizes this complete learning procedure.

\begin{algorithm}[!t]
\caption{Online Learning of GB-SMCF Hyperparameters}
\label{alg:cov_design}
\begin{algorithmic}[1]
\REQUIRE
$\mathcal{X}$;
$\mathbf{z}_{\textrm{aug}}\in\R^{2P}$;
$\sigma_{\textrm{obs}}$;
initial $\theta^{(0)}$;
box constraints $\Theta$.
\ENSURE
$\mathbf{C}_\theta$;
learned optimization variable $\theta$.
\STATE Set $\boldsymbol{\Sigma}_{\textrm{n}}=(\sigma_{\textrm{obs}}^2/2)\I_{2P}$ and initialize $\theta\leftarrow\theta^{(0)}$.
\REPEAT
    \STATE Compute the complex side covariance matrices $\mathbf{K}_{\textrm{r}}\in\C^{P\times P}$ and $\mathbf{K}_{\textrm{t}}\in\C^{P\times P}$ using~\eqref{eq:K-rx}--\eqref{eq:K-tx}, with $[\mathbf{K}_{\textrm{r}}]_{ab}=k_{\textrm{r}}(i_a,i_b)$, $[\mathbf{K}_{\textrm{t}}]_{ab}=k_{\textrm{t}}(j_a,j_b)$, and $x_a=(i_a,j_a)$, $x_b=(i_b,j_b)\in\mathcal{X}$.
    \STATE Form $\mathbf{K}_{\textrm{GB}}(\mathcal{X},\mathcal{X})=\alpha(\mathbf{K}_{\textrm{r}}\circ\mathbf{K}_{\textrm{t}})$ using~\eqref{eq:GB-SM-function}.
    \STATE Form $\widetilde{\mathbf{K}}_{\mathcal{X}\mathcal{X}}=\mathcal{T}\!\big(\mathbf{K}_{\textrm{GB}}(\mathcal{X},\mathcal{X})\big)$ and $\mathbf{C}_\theta=\widetilde{\mathbf{K}}_{\mathcal{X}\mathcal{X}}+\boldsymbol{\Sigma}_{\textrm{n}}$.
    \STATE Compute $\mathcal{L}(\theta)$ via~\eqref{eq:LML} and its gradient via~\eqref{eq:GP-grad-master}.
    \STATE Update $\theta$ within $\Theta$.
\UNTIL{convergence}
\RETURN $\mathbf{C}_\theta,\theta$.
\end{algorithmic}
\end{algorithm}

%%%%%%%%%%%%%%%%%%%%%%%%%%%%%%%%%%%%%%%%%%%%%%%%%%%%%%%%%%%%%%%%%%%%%%%%%%%%%%%%%%%%%%%%%%
\vspace{-2mm}
\subsection{Proposed GB-SMCF-Based GPR Channel Estimator}
\label{SubSec:GPR}
We infer the unobserved channel coefficients on $\mathcal{X}_\star$ by combining the real-coordinate observation model in~\eqref{eq:z_icm} with the zero-mean proper complex GP prior evaluated at the learned hyperparameters. The prior covariance function $k_{\textrm{GB}}(\cdot,\cdot)$ induces the covariance matrices $\mathbf{K}_{\textrm{GB}}(\mathcal{X},\mathcal{X})$, $\mathbf{K}_{\textrm{GB}}(\mathcal{X},\mathcal{X}_\star)$, $\mathbf{K}_{\textrm{GB}}(\mathcal{X}_\star,\mathcal{X})$, and $\mathbf{K}_{\textrm{GB}}(\mathcal{X}_\star,\mathcal{X}_\star)$ on the corresponding index sets through~\eqref{eq:GB-SM-function}. Their real-coordinate counterparts are obtained through $\mathcal{T}(\cdot)$, namely, $\widetilde{\mathbf{K}}_{\mathcal{X}\mathcal{X}}=\mathcal{T}\!\big(\mathbf{K}_{\textrm{GB}}(\mathcal{X},\mathcal{X})\big)$, $\widetilde{\mathbf{K}}_{\mathcal{X}\mathcal{X}_\star}=\mathcal{T}\!\big(\mathbf{K}_{\textrm{GB}}(\mathcal{X},\mathcal{X}_\star)\big)$, $\widetilde{\mathbf{K}}_{\mathcal{X}_\star\mathcal{X}}=\mathcal{T}\!\big(\mathbf{K}_{\textrm{GB}}(\mathcal{X}_\star,\mathcal{X})\big)$, and $\widetilde{\mathbf{K}}_{\mathcal{X}_\star\mathcal{X}_\star}=\mathcal{T}\!\big(\mathbf{K}_{\textrm{GB}}(\mathcal{X}_\star,\mathcal{X}_\star)\big)$.

Under the GP prior, the stacked real-coordinate channel vectors on the observed and unobserved index sets satisfy
\begin{equation}
\label{eq:prior_icm}
\begin{bmatrix}
\mathbf{f}_{\mathcal{X}}\\
\mathbf{f}_{\mathcal{X}_\star}
\end{bmatrix}
\sim
\mathcal{N}\!\left(
\mathbf{0},
\begin{bmatrix}
\widetilde{\mathbf{K}}_{\mathcal{X}\mathcal{X}} & \widetilde{\mathbf{K}}_{\mathcal{X}\mathcal{X}_\star}\\
\widetilde{\mathbf{K}}_{\mathcal{X}_\star\mathcal{X}} & \widetilde{\mathbf{K}}_{\mathcal{X}_\star\mathcal{X}_\star}
\end{bmatrix}
\right),
\end{equation}
with $\widetilde{\mathbf{K}}_{\mathcal{X}\mathcal{X}_\star}=\widetilde{\mathbf{K}}_{\mathcal{X}_\star\mathcal{X}}^{\mathsf{T}}$. Combining the prior in~\eqref{eq:prior_icm} with the observation model in~\eqref{eq:z_icm}, the noisy observations and the unobserved coefficients are jointly Gaussian:
\begin{equation}
\label{eq:joint_obs_pred}
\begin{bmatrix}
\mathbf{z}_{\textrm{aug}}\\
\mathbf{f}_{\mathcal{X}_\star}
\end{bmatrix}
\sim
\mathcal{N}\!\left(
\mathbf{0},
\begin{bmatrix}
\mathbf{C}_\theta & \widetilde{\mathbf{K}}_{\mathcal{X}\mathcal{X}_\star}\\
\widetilde{\mathbf{K}}_{\mathcal{X}_\star\mathcal{X}} & \widetilde{\mathbf{K}}_{\mathcal{X}_\star\mathcal{X}_\star}
\end{bmatrix}
\right).
\end{equation}

Conditioning the joint model in~\eqref{eq:joint_obs_pred} on $\mathbf{z}_{\textrm{aug}}$ yields the posterior distribution on the unobserved set: $\mathbf{f}_{\mathcal{X}_\star}\mid \mathbf{z}_{\textrm{aug}}\sim\mathcal{N}(\boldsymbol{m}_{\mathcal{X}_\star},\boldsymbol{\Sigma}_\star)$, with
\begin{align}
\label{eq:ICM-post-mean}
\boldsymbol{m}_{\mathcal{X}_\star}
&=
\widetilde{\mathbf{K}}_{\mathcal{X}_\star\mathcal{X}}\mathbf{C}_\theta^{-1}\mathbf{z}_{\textrm{aug}},\\
\label{eq:ICM-post-cov}
\boldsymbol{\Sigma}_\star
&=
\widetilde{\mathbf{K}}_{\mathcal{X}_\star\mathcal{X}_\star}
-
\widetilde{\mathbf{K}}_{\mathcal{X}_\star\mathcal{X}}\mathbf{C}_\theta^{-1}\widetilde{\mathbf{K}}_{\mathcal{X}_\star\mathcal{X}}^{\mathsf{T}}.
\end{align}
The corresponding posterior-mean smoother on the observed set is
\begin{equation}
\label{eq:ICM-post-mean-X}
\boldsymbol{m}_{\mathcal{X}}
=
\widetilde{\mathbf{K}}_{\mathcal{X}\mathcal{X}}\mathbf{C}_\theta^{-1}\mathbf{z}_{\textrm{aug}}.
\end{equation}
Thus, $\boldsymbol{m}_{\mathcal{X}}$ and $\boldsymbol{m}_{\mathcal{X}_\star}$ are the posterior means of the stacked real and imaginary channel components on $\mathcal{X}$ and $\mathcal{X}_\star$, respectively, while $\diag(\boldsymbol{\Sigma}_\star)$ quantifies the predictive uncertainty on $\mathcal{X}_\star$.

Writing $\boldsymbol{m}_{\mathcal{X}}=[(\boldsymbol{m}_{\mathcal{X}}^{\Re})^{\mathsf{T}}\ (\boldsymbol{m}_{\mathcal{X}}^{\Im})^{\mathsf{T}}]^{\mathsf{T}}$ and $\boldsymbol{m}_{\mathcal{X}_\star}=[(\boldsymbol{m}_{\mathcal{X}_\star}^{\Re})^{\mathsf{T}}\ (\boldsymbol{m}_{\mathcal{X}_\star}^{\Im})^{\mathsf{T}}]^{\mathsf{T}}$, the corresponding complex posterior-mean estimates are $\widehat{\mathbf{h}}_{\mathcal{X}}=\boldsymbol{m}_{\mathcal{X}}^{\Re}+\mathrm{j}\,\boldsymbol{m}_{\mathcal{X}}^{\Im}\in\C^{P}$ and $\widehat{\mathbf{h}}_{\mathcal{X}_\star}=\boldsymbol{m}_{\mathcal{X}_\star}^{\Re}+\mathrm{j}\,\boldsymbol{m}_{\mathcal{X}_\star}^{\Im}\in\C^{P_\star}$. Combining posterior-mean smoothing on $\mathcal{X}$ with posterior prediction on $\mathcal{X}_\star$ yields the full CSI estimate
\begin{equation}
\label{eq:ReconChannelMtrx}
\widehat{\mathbf{H}}_{\textrm{GPR}}
=
\big[\hat h(i,j)\big]_{i=1,\dots,N_{\textrm{r}}}^{j=1,\dots,N_{\textrm{t}}}
\in\C^{N_{\textrm{r}}\times N_{\textrm{t}}},
\end{equation}
where $\hat h(i,j)$ is taken from $\widehat{\mathbf{h}}_{\mathcal{X}}$ for $(i,j)\in\mathcal{X}$ and from $\widehat{\mathbf{h}}_{\mathcal{X}_\star}$ for $(i,j)\in\mathcal{X}_\star$. Algorithm~\ref{alg:proposed_estimator} summarizes the proposed GB-SMCF-based GPR estimator.

\begin{algorithm}[!t]
\caption{Proposed GB-SMCF-Based GPR Estimator}
\label{alg:proposed_estimator}
\begin{algorithmic}[1]
\REQUIRE
$\Omega_{\textrm{t}}\subset\{1,\dots,N_{\textrm{t}}\}$ with $|\Omega_{\textrm{t}}|=n_{\textrm{t}}$;
$\mathbf{Z}\in\C^{N_{\textrm{r}}\times n_{\textrm{t}}}$ from~\eqref{eq:Zdef};
$\sigma_{\textrm{obs}}$;
initial $\theta^{(0)}$;
box constraints $\Theta$.
\ENSURE
$\widehat{\mathbf{H}}_{\textrm{GPR}}\in\C^{N_{\textrm{r}}\times N_{\textrm{t}}}$ and
$\boldsymbol{\Sigma}_\star$.
\STATE Form $\mathcal{X}$ via~\eqref{eq:X_train_pf} and $\mathcal{X}_\star$ via~\eqref{eq:Xstar_pf}.
\STATE Form $\mathbf{z}=\vecc(\mathbf{Z})$ and $\mathbf{z}_{\textrm{aug}}$ via~\eqref{eq:z_icm}.
\STATE Invoke Algorithm~\ref{alg:cov_design} with $(\mathcal{X},\mathbf{z}_{\textrm{aug}},\sigma_{\textrm{obs}},\theta^{(0)},\Theta)$ to obtain the learned optimization variable $\theta$.
\STATE Evaluate $\mathbf{K}_{\textrm{GB}}(\mathcal{X},\mathcal{X})$, $\mathbf{K}_{\textrm{GB}}(\mathcal{X}_\star,\mathcal{X})$, and $\mathbf{K}_{\textrm{GB}}(\mathcal{X}_\star,\mathcal{X}_\star)$ using the learned model hyperparameters $\vartheta(\theta)$.
\STATE Form the covariance matrices $\widetilde{\mathbf{K}}_{\mathcal{X}\mathcal{X}}$, $\widetilde{\mathbf{K}}_{\mathcal{X}_\star\mathcal{X}}$, and $\widetilde{\mathbf{K}}_{\mathcal{X}_\star\mathcal{X}_\star}$.
\STATE Form $\mathbf{C}_\theta=\widetilde{\mathbf{K}}_{\mathcal{X}\mathcal{X}}+\frac{\sigma_{\textrm{obs}}^2}{2}\I_{2P}$.
\STATE Solve the linear system $\mathbf{C}_\theta\mathbf{q}_\theta=\mathbf{z}_{\textrm{aug}}$, and compute $\boldsymbol{m}_{\mathcal{X}}=\widetilde{\mathbf{K}}_{\mathcal{X}\mathcal{X}}\mathbf{q}_\theta$ and $\boldsymbol{m}_{\mathcal{X}_\star}=\widetilde{\mathbf{K}}_{\mathcal{X}_\star\mathcal{X}}\mathbf{q}_\theta$.
\STATE Form $\widehat{\mathbf{h}}_{\mathcal{X}}=\boldsymbol{m}_{\mathcal{X}}^{\Re}+\mathrm{j}\,\boldsymbol{m}_{\mathcal{X}}^{\Im}$ and $\widehat{\mathbf{h}}_{\mathcal{X}_\star}=\boldsymbol{m}_{\mathcal{X}_\star}^{\Re}+\mathrm{j}\,\boldsymbol{m}_{\mathcal{X}_\star}^{\Im}$, and assemble $\widehat{\mathbf{H}}_{\textrm{GPR}}$ by placing $\widehat{\mathbf{h}}_{\mathcal{X}}$ on $\mathcal{X}$ and $\widehat{\mathbf{h}}_{\mathcal{X}_\star}$ on $\mathcal{X}_\star$.
\STATE Compute $\boldsymbol{\Sigma}_\star=\widetilde{\mathbf{K}}_{\mathcal{X}_\star\mathcal{X}_\star}-\widetilde{\mathbf{K}}_{\mathcal{X}_\star\mathcal{X}}\mathbf{V}$, where $\mathbf{V}$ solves $\mathbf{C}_\theta\mathbf{V}=\widetilde{\mathbf{K}}_{\mathcal{X}_\star\mathcal{X}}^{\mathsf{T}}$.
\RETURN $\widehat{\mathbf{H}}_{\textrm{GPR}}$ and $\boldsymbol{\Sigma}_\star$.
\end{algorithmic}

\end{algorithm}

%%%%%%%%%%%%%%%%%%%%%%%%%%%%%%%%%%%%%%%%%%%%%%%%%%%%%%%%%%%%%%%%%%%%%%%%%%%%%%%%%%%%%%%%%%
%%%%%%%%%%%%%%%%%%%%%%%%%%%%%%%%%%%%%%%%%%%%%%%%%%%%%%%%%%%%%%%%%%%%%%%%%%%%%%%%%%%%%%%%%%
\vspace{-2mm}
\section{Optimization and Complexity Analysis}
\label{Sec:opt_complexity}
The proposed GB-SMCF-based GPR combines online hyperparameter learning with exact GP posterior evaluation. Accordingly, beyond the estimator definition itself, this section characterizes three practical aspects of the proposed framework: the optimization behavior of the objective function, the computational complexity, and the implications of this complexity for larger antenna arrays.

%%%%%%%%%%%%%%%%%%%%%%%%%%%%%%%%%%%%%%%%%%%%%%%%%%%%%%%%%%%%%%%%%%%%%%%%%%%%%%%%%%%%%%%%%%
\vspace{-2mm}
\subsection{Optimization Behavior of the Objective Function}
The GB-SMCF hyperparameters are learned by maximizing the nonconvex marginal-likelihood objective in~\eqref{eq:opt-problem}. Although performed numerically, its local geometry reveals whether the kernel parameters are stably identified from the pilot observations and how sensitive the optimization is to initialization. To illustrate this behavior, Fig.~\ref{fig:convergence} shows a representative view of the objective around a reference solution $\hat{\theta}$.

The left panel of Fig.~\ref{fig:convergence} plots a conditional two-dimensional slice of the marginal-likelihood gap $\Delta\mathcal{L}(\theta)=\mathcal{L}(\hat{\theta})-\mathcal{L}(\theta)$ for a selected pair of transmit-side mean spatial frequencies, while all remaining optimization variables are fixed at $\hat{\theta}$. In this figure, the slice is taken with respect to the dominant transmit spectral component $(\mu_{p^\star,y}^{(\textrm{t})},\mu_{p^\star,z}^{(\textrm{t})})$ with $p^\star=2$. The overlaid curves are the projections of full-dimensional bound-constrained optimizer iterates from six different initializations~\cite{optimizer}, where $S_i$ and $E_i$ denote the start and end of run $i$, respectively. The contour geometry around $\hat{\theta}$ exhibits a smooth anisotropic basin rather than an irregular or fragmented region, indicating that the objective is locally well behaved in these coordinates. The tilted contours show coupling between $\mu_{p^\star,y}^{(\textrm{t})}$ and $\mu_{p^\star,z}^{(\textrm{t})}$, while the unequal principal curvatures indicate different local sensitivities along varying linear combinations of these coordinates.

The right panel of Fig.~\ref{fig:convergence} reports the convergence gap $\mathcal{L}(\hat{\theta})-\mathcal{L}(\theta^{(k)})$ versus iteration for the same multistart runs. The objective gaps of trajectories that enter the dominant basin decay rapidly and terminate at nearly identical objective values, indicating reduced sensitivity to initialization within that basin. Some runs, however, terminate at larger gaps, showing that the full optimization problem in $\theta$ remains nonconvex and initialization dependent. Thus, Fig.~\ref{fig:convergence} supports the following conclusions: i) the kernel-learning objective is locally smooth in the relevant basin, though not globally simple; ii) the bound-constrained optimizer converges reliably once initialized within or near that basin; iii) multistart optimization is useful for robust training; and iv) the selected transmit-side frequencies exhibit nontrivial local coupling, with one local direction more sharply constrained by the data than another.

\begin{figure}[t]
    \centering
    \input{Figures/contour_init_sensitivity}
    \caption{Multistart marginal-likelihood optimization behavior for the proposed GB-SMCF-based GPR model.}
    \label{fig:convergence}
\end{figure}

\begin{table}[t]
\centering
\caption{Dominant estimator-side computational complexity of different estimators.}
\label{tab:complexity}
\renewcommand{\arraystretch}{1.15}
\setlength{\tabcolsep}{3.5pt}
\scriptsize
\begin{tabular}{|p{1.95cm}|p{1.8cm}|p{4.25cm}|}
\hline
\textbf{Algorithm}
& \textbf{Complexity}
& \textbf{Remarks} \\
\hline
\textbf{Proposed GB-SMCF-based GPR}
& $\mathcal{O}\!\left(I_{\textrm{opt}}P^3\right)$
& Dominant hyperparameter-learning cost for the exact real-augmented GP, with $P=N_{\textrm{r}}n_{\textrm{t}}$. The covariance matrix has size $2P\times 2P$. After the final factorization, posterior-mean evaluation on $P_\star$ test points costs $\mathcal{O}(P P_\star)$; explicitly forming the full $\boldsymbol{\Sigma}_\star$ costs $\mathcal{O}(P^2P_\star+P P_\star^2)$. \\
\hline
\textbf{LS}
& $\mathcal{O}\!\left(N_{\textrm{r}}N_{\textrm{t}}\right)$
& Estimator-side cost after the common preprocessing step used to form $\mathbf{Z}$; under full-array training, $\widehat{\mathbf{H}}_{\textrm{LS}}=\mathbf{Z}$. \\
\hline
\textbf{Isotropic linear MMSE}~\cite{demir2022channel}
& $\mathcal{O}\!\left((N_{\textrm{r}}N_{\textrm{t}})^3\right)$
& Dense fixed-covariance linear MMSE (LMMSE) with a system of size $N_{\textrm{r}}N_{\textrm{t}}$; the error covariance is determined by the assumed prior and is not learned online from the current block. \\
\hline
\textbf{OMP}~\cite{do2008sparsity}
& $\mathcal{O}\!\left(I_{\textrm{OMP}} M G\right)$
& Coarse leading-order correlation cost for $M$ measurements and dictionary size $G$; additional least-squares and support-update costs depend on the sparsity level. Under full-array training, $M=N_{\textrm{r}}N_{\textrm{t}}$. \\
\hline
\textbf{AMP}~\cite{donoho2010message}
& $\mathcal{O}\!\left(I_{\textrm{AMP}} M G\right)$
& Coarse leading-order cost for dense matrix-vector iterations with $M$ measurements and dictionary size $G$. Per-iteration cost is low, but performance depends on the sensing model, sparsity assumptions, and implementation. Under full-array training, $M=N_{\textrm{r}}N_{\textrm{t}}$. \\
\hline
\end{tabular}
\end{table}
%%%%%%%%%%%%%%%%%%%%%%%%%%%%%%%%%%%%%%%%%%%%%%%%%%%%%%%%%%%%%%%%%%%%%%%%%%%%%%%%%%%%%%%%%%
\vspace{-2mm}
\subsection{Computational Complexity}
To keep the comparison scope consistent across different estimation methods, Table~\ref{tab:complexity} reports dominant complexity after the common pilot-correlation preprocessing used to form $\mathbf{Z}$. For the proposed GB-SMCF-based GPR method, hyperparameter learning is dominated by repeated factorizations of $\mathbf{C}_\theta\in\R^{2P\times 2P}$. One evaluation of the log marginal likelihood and its gradient uses one Cholesky factorization of $\mathbf{C}_\theta$, together with lower-order triangular solves, log-determinant evaluation, and derivative contractions. Thus, each optimization step costs $\approx\mathcal{O}(P^3)$ time and $\approx\mathcal{O}(P^2)$ memory, and the dominant training cost over $I_{\textrm{opt}}$ iterations is $\mathcal{O}(I_{\textrm{opt}}P^3)$. After the final factorization has been computed, posterior-mean evaluation on $P_\star=|\mathcal{X}_\star|$ test points costs $\mathcal{O}(P P_\star)$. If the full posterior covariance $\boldsymbol{\Sigma}_\star\in\R^{2P_\star\times 2P_\star}$ is formed explicitly, the additional cost is $\mathcal{O}(P^2P_\star+P P_\star^2)$ time and $\mathcal{O}(P_\star^2)$ memory; if only marginal predictive variances are required, the covariance-evaluation cost reduces accordingly.

The separable GB-SMCF contributes only a lower-order kernel-assembly cost. For a general training set, $\mathbf{K}_{\textrm{GB}}(\mathcal{X},\mathcal{X})=\alpha(\mathbf{K}_{\textrm{r}}\circ\mathbf{K}_{\textrm{t}})$, so assembling the side Gram matrices and their derivative contributions costs $\mathcal{O}(D_\theta P^2)$, where $D_\theta$ is the number of optimized hyperparameters. When the number of parameters per mixture component is treated as constant, this reduces to $\mathcal{O}((Q_{\textrm{r}}+Q_{\textrm{t}})P^2)$, which is dominated by the cubic factorization term. When the training set is Cartesian, i.e., $\mathcal{X}=\{1,\cdots, N_{\textrm{r}}\}\times\Omega_{\textrm{t}}$ with indices ordered lexicographically, the complex covariance matrix admits the exact Kronecker form $\mathbf{K}_{\textrm{GB}}(\mathcal{X},\mathcal{X})=\alpha\big(\mathbf{K}_{\textrm{r}}^{\textrm{lat}}\otimes\mathbf{K}_{\textrm{t}}^{\Omega_{\textrm{t}}}\big)$, where $\mathbf{K}_{\textrm{r}}^{\textrm{lat}}$ is the receive-side lattice Gram matrix and $\mathbf{K}_{\textrm{t}}^{\Omega_{\textrm{t}}}$ is the transmit-side Gram matrix restricted to the active subset $\Omega_{\textrm{t}}$. Structured implementations can exploit this Kronecker form most naturally in the equivalent complex Hermitian system; the real-augmented covariance $\widetilde{\mathbf{K}}_{\mathcal{X}\mathcal{X}}=\mathcal{T}\!\big(\mathbf{K}_{\textrm{GB}}(\mathcal{X},\mathcal{X})\big)$ preserves algebraic equivalence but is not, in general, itself a plain Kronecker product. Thus, unless such structure is explicitly exploited in the linear algebra, the exact implementation considered here retains cubic worst-case training complexity. Finally, the above discussion is asymptotic and does not by itself establish real-time operation: supporting online deployment requires comparing the measured latency and memory footprint of both hyperparameter learning and posterior evaluation with the coherence-time budget of the target system~\cite{hensman2015scalable, gardner2018gpytorch}.
%%%%%%%%%%%%%%%%%%%%%%%%%%%%%%%%%%%%%%%%%%%%%%%%%%%%%%%%%%%%%%%%%%%%%%%%%%%%%%%%%%%%%%%%%%
\vspace{-2mm}
\subsection{Scalability for Large Arrays}
For larger arrays, i.e., larger $N_{\textrm{r}}$ or $n_{\textrm{t}}$ and hence larger $P=N_{\textrm{r}}n_{\textrm{t}}$, dense-Cholesky exact GP training rapidly becomes the computational bottleneck. Standard exact GP regression scales as $\mathcal{O}(P^3)$ in time and $\mathcal{O}(P^2)$ in memory, so in practice dense exact methods are restricted to moderate training sizes, e.g., $P\lesssim 10^4$, unless additional structure is exploited~\cite{hensman2015scalable}. In the present setting, this limitation applies directly to the exact real-augmented implementation above, with only a constant-factor difference relative to the equivalent complex Hermitian formulation. Several principled extensions can improve scalability:
\begin{enumerate}
    \item \emph{Matrix-vector-based exact solvers}, such as blackbox matrix--matrix GP, replace dense factorizations with iterative matrix-vector operations, reducing the effective cost of exact GP training and enabling GPU-accelerated inference on large data sets~\cite{gardner2018gpytorch};
    \item \emph{Inducing-point and Nystr\"om approximations} replace the full kernel with a rank-$M$ approximation, yielding the standard sparse-GP complexity $\mathcal{O}(P M^2+M^3)$ with memory $\mathcal{O}(P M+M^2)$ for $M\ll P$~\cite{hensman2015scalable};
    \item \emph{Structured kernel interpolation} yields near-linear complexity; in conjunction with Kronecker structure, it achieves $\mathcal{O}(P+D m^{1+1/D})$ computations and $\mathcal{O}(P+D m^{2/D})$ storage for $D$-dimensional inputs and $m$ inducing-grid points~\cite{wilson2015kernel}; and
    \item \emph{Exact Kronecker methods}, for Cartesian grids and tensor-product kernels, evaluate linear solves and log-determinants from factor eigendecompositions rather than from a dense factorization~\cite{lin2025scalable}.
\end{enumerate}

\noindent In the proposed model, the Cartesian training set $\mathcal{X}=\{1,\cdots,N_{\textrm{r}}\}\times\Omega_{\textrm{t}}$ and the separable GB-SMCF are directly compatible with such structured exact or approximate GP solvers. Thus, the exact method considered in this paper is best suited to moderate values of $P$, whereas deployment for larger $N_{\textrm{r}}$ or $n_{\textrm{t}}$ requires either Kronecker-exploiting exact solvers or sparse approximations~\cite{gardner2018gpytorch,lin2025scalable,hensman2015scalable,wilson2015kernel}. These large-scale approximations will be considered in future work, since the goal of this work is to introduce the proposed GPR-based CSI acquisition framework and its GB-SMCF kernel.

\begin{table}[t]

\centering
\caption{Simulation parameters and optimizer settings.}
\label{tab:sim_params}
\scriptsize
\renewcommand{\arraystretch}{1.1}
\setlength{\tabcolsep}{4pt}
\begin{tabular}{|p{0.39\columnwidth}|p{0.13\columnwidth}|p{0.38\columnwidth}|}
\hline
\textbf{Parameter} & \textbf{Symbol} & \textbf{Value} \\
\hline
\multicolumn{3}{|c|}{\textbf{System and training}} \\
\hline
Receive $\times$ transmit antennas
& $N_{\textrm{r}}\times N_{\textrm{t}}$
& $16\times 16$ \\
Element spacing
& $d/\lambda_c$
& $0.5$ \\
Carrier frequency
& $f_c$
& $28~\mathrm{GHz}$ \\
Coherence block length
& $T_c$
& $100$ \\
Per-active-antenna pilot power
& $P_{\textrm{A}}$
& $1$ \\
Pilot length
& $T$
& $n_{\textrm{t}}$ \\
Pilot normalization
& $\mathbf{S}\mathbf{S}^{\Hsf}$
& $T\I_{n_{\textrm{t}}}$ \\
Training SNR
& $\mathrm{SNR}$
& %$TP_{\textrm{A}}/\sigma_n^2$
$1/\sigma_{\textrm{obs}}^2$ \\
Total training energy
& $E_{\textrm{tr}}$
& $n_{\textrm{t}}TP_{\textrm{A}}=n_{\textrm{t}}^2P_{\textrm{A}}$ \\
Monte Carlo runs
& --
& $100$ \\
\hline
\multicolumn{3}{|c|}{\textbf{Saleh--Valenzuela channel model parameters}} \\
\hline
Number of clusters & $C$ & $3$ \\
Rays per cluster & $R_c$ & $2$ \\
Angular spreads & $\sigma_\phi,\sigma_\psi$ & $10^\circ,\,7^\circ$ \\
\hline
\multicolumn{3}{|c|}{\textbf{3GPP TR~38.901 CDL-A}} \\
\hline
Scenario
& --
& Indoor office \\
Delay profile
& --
& CDL-A \\
Delay-spread scaling
& --
& $20~\mathrm{ns}$ at $28~\mathrm{GHz}$ \\
Angular/delay parameters
& --
& As per TR~38.901 CDL-A~\cite{3GPP_TR_38_901} \\
\hline
\multicolumn{3}{|c|}{\textbf{GB-SMCF model and optimizer settings}} \\
\hline
Kernel variance scale
& $\alpha$
& $[10^{-3},\,3\times10^{1}]$ \\
Receive mixture order
& $Q_{\textrm{r}}$
& $3$ \\
Transmit mixture order
& $Q_{\textrm{t}}$
& $3$ \\
Receive mixture weights
& $\{\pi_q^{(\textrm{r})}\}_{q=1}^{Q_{\textrm{r}}}$
& $\pi_q^{(\textrm{r})}\ge 0$, $\sum_q \pi_q^{(\textrm{r})}=1$\\%; enforced by softmax reparameterization \\

Transmit mixture weights
& $\{\pi_p^{(\textrm{t})}\}_{p=1}^{Q_{\textrm{t}}}$
& $\pi_p^{(\textrm{t})}\ge 0$, $\sum_p \pi_p^{(\textrm{t})}=1$\\%; enforced by softmax reparameterization \\
Receive spatial frequencies
& $\mu_{q,y}^{(\textrm{r})},\,\mu_{q,z}^{(\textrm{r})}$
& $[-0.5,\,0.5]$ \\
Transmit spatial frequencies
& $\mu_{p,y}^{(\textrm{t})},\,\mu_{p,z}^{(\textrm{t})}$
& $[-0.5,\,0.5]$ \\
Receive spectral variances
& $v_{q,y}^{(\textrm{r})},\,v_{q,z}^{(\textrm{r})}$
& $[6\times10^{-4},\,10^{-1}]$ \\
Transmit spectral variances
& $v_{p,y}^{(\textrm{t})},\,v_{p,z}^{(\textrm{t})}$
& $[6\times10^{-4},\,10^{-1}]$ \\
%Optimization domain & $\Theta$ & Feasible box-constrained set for $\theta$ \\
Optimizer
& --
& Bound-constrained optimizer~\cite{optimizer} \\
Multistart initialization
& --
& $6$ starts \\
\hline
\multicolumn{3}{|c|}{\textbf{Numerical stability}} \\
\hline
Kernel jitter\tablefootnote{The kernel jitter is added to the diagonal of the covariance matrix to improve numerical stability during the Cholesky decomposition.}
& $\epsilon$
& $10^{-8}$ added to $\mathbf{C}_\theta$ \\
\hline
\end{tabular}
\end{table}
%%%%%%%%%%%%%%%%%%%%%%%%%%%%%%%%%%%%%%%%%%%%%%%%%%%%%%%%%%%%%%%%%%%%%%%%%%%%%%%%%%%%%%%%%%
%%%%%%%%%%%%%%%%%%%%%%%%%%%%%%%%%%%%%%%%%%%%%%%%%%%%%%%%%%%%%%%%%%%%%%%%%%%%%%%%%%%%%%%%%%
\vspace{-3mm}
\section{Simulation Results}
\label{sec:simulation_results}
We evaluate the proposed GB-SMCF-based GPR estimator under two channel models: i) the Saleh--Valenzuela model adopted from~\cite{SV1987channelModel}, and ii) the Third Generation Partnership Project (3GPP) Technical Report (TR)~38.901 clustered delay line (CDL) model with the CDL-A delay profile for the indoor-office scenario adopted from~\cite{3GPP_TR_38_901}. Unless stated otherwise, Table~\ref{tab:sim_params} lists the simulation parameters and optimizer settings. 
To enhance spatial coverage, the active set $\Omega_{\textrm{t}}$ is selected by the equispaced rule~\cite{shah2026improved,shah2025lowoverhead}
\begin{equation}
\resizebox{\columnwidth}{!}{$
    \Omega_{\textrm{t}} =
\begin{cases}
\{1\}, & n_{\textrm{t}} = 1,\\[4pt]
\left\{1 + \operatorname{round}\left(\dfrac{(m-1)(N_{\textrm{t}}-1)}{n_{\textrm{t}}-1}\right) : m = 1,\dots,n_{\textrm{t}} \right\}, & n_{\textrm{t}} \ge 2,
\end{cases}$}
\end{equation}
followed, if necessary, by a deterministic tie-breaking completion step to ensure $|\Omega_{\textrm{t}}|= n_{\textrm{t}}$. The proposed GB-SMCF-based GPR estimator can accommodate arbitrary transmit-antenna selection patterns; the reported results use the above equispaced rule.

\vspace{-1mm}
\paragraph{Channel models}
Under the Saleh--Valenzuela channel model~\cite{SV1987channelModel}, the channel matrix is generated as
\begin{equation}
\label{eq:SV}
\mathbf{H}
=
\sum_{c=1}^{C}\sum_{r=1}^{R_c}
\beta_{c,r}\,
\mathbf{a}_{\textrm{r}}(\boldsymbol{\phi}_{c,r})\,
\mathbf{a}_{\textrm{t}}^{\Hsf}(\boldsymbol{\psi}_{c,r})
\in \C^{N_{\textrm{r}}\times N_{\textrm{t}}},
\end{equation}
where $C$ is the number of clusters, $R_c$ is the number of rays in cluster $c$, $\beta_{c,r}\in\C$ is the complex gain of ray $(c,r)$, and $\mathbf{a}_{\textrm{t}}(\cdot)$ and $\mathbf{a}_{\textrm{r}}(\cdot)$ denote the transmit and receive URA steering vectors, respectively. The Saleh--Valenzuela parameters used in the simulations are listed in Table~\ref{tab:sim_params}, unless stated otherwise. For the 3GPP benchmark, channels are generated according to the indoor-office CDL-A profile in TR~38.901~\cite{3GPP_TR_38_901}; a single narrowband channel realization at $f_c$ is extracted after delay-spread scaling to $20~\mathrm{ns}$, with all remaining CDL-A parameters taken directly from the standard specification. Both channel generators are external to the estimator design and therefore introduce no systematic bias in favor of the proposed GB-SMCF-based GPR framework~\cite{shah2026improved,shah2025lowoverhead}.

\vspace{-1mm}
\paragraph{Baseline channel estimators}
Unless stated otherwise, the LS, isotropic LMMSE, OMP, and AMP baselines are evaluated under full-array training configuration, i.e., $n_{\textrm{t}}=N_{\textrm{t}}$. These comparisons therefore emphasize training-efficiency trade-offs rather than equal pilot budget estimator performance. In contrast, the SqExp- and Mat\'ern-kernel GPR baselines are evaluated with the same pilot budgets as the proposed method, since they are also GP-based predictors on partial-array observations. All compared methods use the same channel realizations, active-antenna subsets, and noise realizations in each Monte Carlo run. The compared estimators are:
i) \textit{LS:} under full-array training, $\widehat{\mathbf{H}}_{\textrm{LS}}=\mathbf{Z}$;
ii) \textit{isotropic LMMSE}~\cite{demir2022channel}: the isotropic prior covariance $\mathbf{R}_{\textrm{iso}}\in\C^{N_{\textrm{r}}N_{\textrm{t}}\times N_{\textrm{r}}N_{\textrm{t}}}$ has entries $[\mathbf{R}_{\textrm{iso}}]_{mn}
= \sigma_h^2\, \sinc\!\Big(\frac{2\pi}{\lambda_c}\|\mathbf{x}_m-\mathbf{x}_n\|_2\Big)$, where $\sigma_h^2=1$, $\sinc(x)=\sin(x)/x$, and $\mathbf{x}_n$ denotes the physical coordinate associated with the $n$-th entry of $\mathbf{h}$~\cite{demir2022channel,Jieao2025IEEE_TIT}. With the sampling matrix $\boldsymbol{\Phi}=(\mathbf{F}^{\mathsf{T}}\otimes \I_{N_{\textrm{r}}})$ from~\eqref{eq:vecModel}, the estimator is
$\widehat{\mathbf{h}}_{\textrm{LMMSE}}
=
\mathbf{R}_{\textrm{iso}}\boldsymbol{\Phi}^{\Hsf}
\Big(\boldsymbol{\Phi}\mathbf{R}_{\textrm{iso}}\boldsymbol{\Phi}^{\Hsf}
+\sigma_{\textrm{obs}}^2\I_{N_{\textrm{r}}n_{\textrm{t}}}\Big)^{-1}\mathbf{z}$;
iii) \textit{standard GPR kernels:} SqExp and Mat\'ern kernels~\cite{rasmussen2006gaussian}, implemented under the same proper complex GP and exact real-augmented inference framework as the proposed method, with identical online marginal-likelihood optimization and Monte Carlo protocols. Their kernel forms and implementation details are adopted from~\cite{rasmussen2006gaussian,shah2025lowoverhead,shah2026improved};
iv) \textit{sparsity-based estimators:} OMP~\cite{do2008sparsity} and AMP~\cite{donoho2010message}, implemented with the same baseline dictionary construction and numerical settings as in~\cite{shah2026improved,shah2025lowoverhead}. AMP uses shrinkage parameter $\tau_{\textrm{AMP}}=1.2$, while OMP uses a fixed sparsity level $K=7$ in the corresponding experiment. The perfect CSI baseline serves to gauge how closely the proposed estimator approaches the almost unachievable ideal  performance.

%%%%%%%%%%%%%%%%%%%%%%%%%%%%%%%%%%%%%%%%%%%%%%%%%%%%%%%%%%%%%%%%%%%%%%%%%%%%%%%%%%%%%%%%%%
\vspace{-2mm}
\subsection{Normalized Mean-Square Error}
\label{subsec:NMSE_results}

We first evaluate the normalized mean-square error (NMSE), $\mathrm{NMSE} = \mathbb E\!\left[
\frac{\|\widehat{\mathbf H}-\mathbf H\|_{\mathrm F}^2}{\|\mathbf H\|_{\mathrm F}^2}
\right]$, where $\widehat{\mathbf H}$ denotes the estimate from the proposed method or a baseline. The expectation is approximated by $100$ Monte Carlo trials over independent channel and noise realizations.

\begin{figure}[t]
    \centering
    \input{Figures/nmse_plot_SV_16}
    \caption{NMSE versus SNR for the proposed GB-SMCF-based GPR estimator and the LS, isotropic LMMSE, OMP, and AMP baselines for a $16\times16$ Saleh--Valenzuela channel model.}
    \label{fig:NMSE_SV_16}
\end{figure}

We begin with the Saleh--Valenzuela channel model. Fig.~\ref{fig:NMSE_SV_16} shows that under reduced pilot budgets, the proposed estimator maintains strong NMSE performance relative to the compared methods. With $n_{\textrm{t}}=8$, corresponding to a $50\%$ reduction in pilot overhead, the proposed method outperforms the full-array training baselines across the considered SNR range. Even with $n_{\textrm{t}}=4$, corresponding to a $75\%$ pilot reduction, it remains competitive, outperforming AMP and OMP at low SNR and LS and isotropic LMMSE over a nontrivial SNR regime.

This behavior is consistent with the structure of the proposed prior. The GB-SMCF explicitly models geometry-dependent receive- and transmit-side correlations through spectral-mixture components, making it better matched to clustered multipath channels than estimators that ignore or simplify the spatial covariance structure. In particular, LS treats the coefficients independently and therefore cannot infer unobserved entries from spatial dependence; isotropic LMMSE uses a simplified covariance model that can be mismatched to the anisotropic correlations induced by array geometry and clustered scattering; and OMP/AMP primarily exploit angular sparsity rather than a calibrated spatial covariance model. Hence, the observed gains suggest that the proposed estimator is not merely denoising the observed coefficients, but is learning a structured spatial prior that enables prediction of unobserved channel coefficients from incomplete antenna-indexed observations.

\begin{figure}[t]
    \centering
    \input{Figures/nmse_plot_SV_GB_SM_Mat_SE_16}
    \caption{NMSE versus SNR for the proposed GB-SMCF-based GPR estimator and GPR baselines with SqExp and Mat\'ern kernels for a $16\times16$ Saleh--Valenzuela channel model.}
    \label{fig:NMSE_SV_Mat_SE_16}
\end{figure}

To isolate the effect of the proposed kernel from the GP framework itself, Fig.~\ref{fig:NMSE_SV_Mat_SE_16} compares the GB-SMCF with GPR baselines using SqExp and Mat\'ern kernels. The proposed GB-SMCF remains superior, including in the reduced-pilot setting $n_{\textrm{t}}=8$, which shows that the gain is not due solely to marginal-likelihood learning but also to the inductive bias encoded by the GB-SMCF. The Mat\'ern kernel performs better than the SqExp kernel, which is plausible because the Mat\'ern prior allows rougher sample paths than the overly smooth SqExp prior~\cite{rasmussen2006gaussian}. However, both remain inferior to the proposed GB-SMCF, indicating that standard distance-based stationary kernels are less effective at representing the structured angular correlation induced by clustered MIMO propagation.

We next assess robustness to model mismatch by moving from the synthetic Saleh--Valenzuela setting to the standardized 3GPP TR~38.901 CDL-A model. As shown in Fig.~\ref{fig:NMSE_CDL_16}, the same qualitative behavior is preserved: under full-array training, i.e., $n_{\textrm{t}}=16$, the proposed estimator achieves the lowest NMSE over the considered SNR range, and with $n_{\textrm{t}}=8$ it continues to outperform the baselines despite a $50\%$ pilot reduction. This suggests that the proposed kernel is not narrowly tied to the Saleh--Valenzuela channel generator and retains useful predictive structure under the standardized CDL-A benchmark. Taken together, Figs.~\ref{fig:NMSE_SV_16} and~\ref{fig:NMSE_CDL_16} indicate that the proposed method is effective under both a physically interpretable clustered model and a standardized benchmark.

\begin{figure}[t]
    \centering
    \input{Figures/nmse_plot_CDL_16}
    \caption{NMSE versus SNR for the proposed GB-SMCF-based GPR estimator and the LS, isotropic LMMSE, OMP, and AMP baselines for a $16\times16$ 3GPP TR~38.901 CDL-A channel model.}
    \label{fig:NMSE_CDL_16}
\end{figure}

\begin{figure}[t]
    \centering
    \input{Figures/nmse_plot_SV_CDL_GB_SM_Mat_SE}
    \caption{NMSE versus SNR for the proposed GB-SMCF, SqExp, and Mat\'ern kernels under the Saleh--Valenzuela and 3GPP TR~38.901 CDL-A channel models for a $16\times16$ MIMO link.}
    \label{fig:NMSE_SV_CDL_kernel_16}
\end{figure}

Finally, Fig.~\ref{fig:NMSE_SV_CDL_kernel_16} compares the kernel-level behavior of the GPR estimators under both the Saleh--Valenzuela and standardized 3GPP TR~38.901 CDL-A channel models for the same $16\times16$ MIMO link. The proposed GB-SMCF achieves the lowest NMSE under the Saleh--Valenzuela model and remains superior under the CDL-A model, demonstrating that its gain is not limited to a single channel generator. This aligns with the GB-SMCF design: spectral-mixture components represent geometry-dependent oscillatory correlations naturally induced by clustered multipath propagation. The stronger performance under the Saleh--Valenzuela model is expected because the Saleh--Valenzuela channel is directly governed by a clustered geometric scattering structure, whereas CDL-A introduces richer standardized angular and delay-domain characteristics that create a more challenging mismatch scenario.

In Fig.~\ref{fig:NMSE_SV_CDL_kernel_16}, the SqExp and Mat\'ern kernels show similar NMSE behavior, with the Mat\'ern kernel slightly outperforming the SqExp kernel. This is plausible because the Mat\'ern prior allows rougher spatial variations than the overly smooth SqExp prior~\cite{rasmussen2006gaussian}. However, both kernels remain distance-based stationary covariance models and do not explicitly encode the angularly oscillatory transmit--receive correlation structure of MIMO channels. The superior performance of the GB-SMCF therefore confirms that the observed gain arises from the proposed physics-aligned covariance structure, rather than from the GP framework alone.

%%%%%%%%%%%%%%%%%%%%%%%%%%%%%%%%%%%%%%%%%%%%%%%%%%%%%%%%%%%%%%%%%%%%%%%%%%%%%%%%%%%%%%%%%%
\vspace{-2mm}
\subsection{Energy--Accuracy Trade-off}
\label{subsec:energy_accuracy_tradeoff}
The NMSE results in Section~\ref{subsec:NMSE_results} show that the proposed estimator remains accurate under reduced pilot budgets. To further examine the trade-off between training energy and estimation accuracy, Fig.~\ref{fig:energy_tradeoff} plots NMSE versus the normalized pilot-budget ratio $n_{\textrm{t}}/N_{\textrm{t}}$ for a $16\times16$ MIMO channel. Since the instantaneous training power equals $n_{\textrm{t}}P_{\textrm{A}}$, this ratio is also the normalized instantaneous training-power ratio. Moreover, because Table~\ref{tab:sim_params} uses $T=n_{\textrm{t}}$, the corresponding normalized total training energy over one training phase is $(n_{\textrm{t}}/N_{\textrm{t}})^2$. Each curve corresponds to a fixed SNR in the range $[-15,10]$ dB, and the vertical guide lines mark the evaluated pilot budgets $n_{\textrm{t}}\in\{2,4,8,16\}$.

\begin{figure}[t]
    \centering
    \input{Figures/energy_tradeoff}
    \caption{Training cost versus accuracy trade-off for the proposed GB-SMCF-based GPR estimator.}
    \label{fig:energy_tradeoff}
\end{figure}

Several conclusions follow. First, for each SNR, NMSE decreases as $n_{\textrm{t}}$ increases over the considered pilot budgets. This is expected: observing more pilot-bearing transmit antennas provides more information for both hyperparameter learning and channel reconstruction. Second, the gain is not uniform across the pilot-budget range. The transition from very sparse training ($n_{\textrm{t}}=2$ or $4$) to moderate training ($n_{\textrm{t}}=8$) yields a relatively large NMSE reduction, whereas the additional gain from $n_{\textrm{t}}=8$ to $n_{\textrm{t}}=16$ is smaller. This indicates diminishing returns once the dominant spatial structure has already been identified. Third, the separation between the SNR-dependent curves becomes more pronounced as the pilot budget increases. Under full or near-full-array training, SNR improvements translate into substantial NMSE reductions because the estimator can exploit the improved observation quality. In contrast, at very small pilot budgets, the benefit of increasing SNR is more limited, since performance is then constrained not only by noise but also by the scarcity of spatial observations. Thus, in the low-budget regime, the dominant bottleneck is measurement insufficiency rather than observation noise alone.

%%%%%%%%%%%%%%%%%%%%%%%%%%%%%%%%%%%%%%%%%%%%%%%%%%%%%%%%%%%%%%%%%%%%%%%%%%%%%%%%%%%%%%%%%%
\vspace{-2mm}
\subsection{Spectral Efficiency}
\label{subsec:SE_results}

To assess how well each estimator preserves the channel information relevant to data detection, we evaluate the SE of a spatially multiplexed Saleh--Valenzuela MIMO channel under a linear receiver designed from the estimated channel~\cite{shah2025lowoverhead}. For a given estimate $\widehat{\mathbf{H}}\in\C^{N_{\textrm{r}}\times N_{\textrm{t}}}$, we use the linear MMSE detector $\mathbf{G}(\widehat{\mathbf{H}})=\big(\widehat{\mathbf{H}}\widehat{\mathbf{H}}^{\Hsf}+\frac{N_{\textrm{t}}}{\rho}\I_{N_{\textrm{r}}}\big)^{-1}\widehat{\mathbf{H}}=[\mathbf{g}_1,\ldots,\mathbf{g}_{N_{\textrm{t}}}]$, where $\rho$ denotes the total data-phase SNR under equal power allocation across the $N_{\textrm{t}}$ streams. Let $\mathbf{h}_k$ and $\mathbf{g}_k$ denote the $k$th columns of the true channel $\mathbf{H}$ and the detector $\mathbf{G}$, respectively. The post-equalization signal-to-interference-plus-noise ratio (SINR) of stream $k$ is
\begin{equation}
\label{eq:sinr_se}
\mathrm{SINR}_k(\widehat{\mathbf{H}})
=
\frac{|\mathbf{g}_k^{\Hsf}\mathbf{h}_k|^2}
{\sum_{j\neq k}|\mathbf{g}_k^{\Hsf}\mathbf{h}_j|^2
+\frac{N_{\textrm{t}}}{\rho}\|\mathbf{g}_k\|^2}.
\end{equation}
The resulting SE is
\begin{equation}
\label{eq:se_metric}
\mathrm{SE}(\widehat{\mathbf{H}})
=
\Big(1-\frac{T}{T_c}\Big)
\sum_{k=1}^{N_{\textrm{t}}}
\log_2\!\bigl(1+\mathrm{SINR}_k(\widehat{\mathbf{H}})\bigr),
\end{equation}
where the pre-log factor accounts for pilot overhead. In the simulations, $T=n_{\textrm{t}}$, so the pre-log becomes $1-n_{\textrm{t}}/T_c$. Thus, SE reflects the fundamental trade-off between channel-estimation quality and training overhead: increasing the pilot budget generally improves the receive filter through a more accurate estimate, but it also reduces the fraction of channel uses available for data transmission. We evaluate the proposed estimator for different pilot budgets together with the LS, isotropic LMMSE, and OMP baselines. Perfect CSI with zero pilot overhead is also included as a genie-aided upper bound.

\begin{figure}[t]
    \centering
    \input{Figures/se_vs_snr}
    \caption{SE versus SNR. The proposed GB-SMCF-based GPR estimator is shown for pilot budgets $n_{\textrm{t}}\in\{4,8,16\}$, corresponding to $75\%$, $50\%$, and $0\%$ reductions in pilot overhead, respectively.}
    \label{fig:SE_for_all}
\end{figure}

Fig.~\ref{fig:SE_for_all} shows this trade-off for a $16\times16$ MIMO channel. The proposed configuration with $n_{\textrm{t}}=4$ achieves the highest SE in the low-to-moderate SNR regime, approximately up to $5$ dB. In this regime, the system is not yet strongly interference-limited, so the degradation caused by fewer pilots remains moderate, whereas the pre-log gain from reduced training overhead is substantial. As the SNR increases, however, the detector becomes more sensitive to directional inaccuracies in $\widehat{\mathbf{H}}$, since residual multi-stream interference rather than thermal noise becomes the dominant limitation. In that regime, the improved channel accuracy obtained with $n_{\textrm{t}}=8$ or $16$ compensates for the larger pilot overhead, and the SE advantage of the smallest pilot budget disappears. A second observation is that the curves for the proposed method with $n_{\textrm{t}}=8$ and $n_{\textrm{t}}=16$ remain close over most of the SNR range. This indicates that, beyond a moderate pilot budget, additional training yields only marginal SINR improvement relative to the associated pre-log penalty. In other words, once the dominant spatial structure has been learned reliably, further pilots provide diminishing returns in SE. This behavior is consistent with the strong spatial prediction capability of the proposed GB-SMCF prior.

Among the baselines, OMP performs better than LS and isotropic LMMSE, consistent with its ability to exploit angular sparsity. Nevertheless, it remains inferior to the proposed GPR-based schemes because support mismatch, basis discretization, and coefficient-estimation error directly degrade the receive filter and hence the post-equalization SINR. LS performs poorly because it imposes no structural prior, making the detector highly sensitive to noise and estimation error. The isotropic LMMSE baseline is more robust at low SNR, where regularization suppresses noise amplification, but it becomes increasingly limited at moderate and high SNR because of covariance mismatch.

%%%%%%%%%%%%%%%%%%%%%%%%%%%%%%%%%%%%%%%%%%%%%%%%%%%%%%%%%%%%%%%%%%%%%%%%%%%%%%%%%%%%%%%%%%
\vspace{-2mm}
\subsection{Critical Discussion}
\label{subsec:critical_discussion}

Table~\ref{tab:comparison_snr0} summarizes representative operating points at ${\rm SNR}=0\,\mathrm{dB}$ for a $16\times16$ MIMO link. The relative SE is normalized by the genie-aided perfect-CSI upper bound. The pilot- and energy-savings figures are measured relative to the full-training case $n_{\textrm{t}}=N_{\textrm{t}}$. Since each active antenna radiates fixed power $P_{\textrm{A}}$ per pilot channel use and $T=n_{\textrm{t}}$, the total training energy scales as $E_{\textrm{tr}}=n_{\textrm{t}}TP_{\textrm{A}}=n_{\textrm{t}}^2P_{\textrm{A}}$. The cost column reports dominant estimator-side complexity after the common pilot-correlation preprocessing used to form $\mathbf{Z}$; for GPR-based methods, it records only the leading hyperparameter-learning term $\mathcal{O}(I_{\textrm{opt}}P^3)$ with $P=N_{\textrm{r}}n_{\textrm{t}}$, excluding lower-order posterior evaluation and explicit formation of the full $\boldsymbol{\Sigma}_\star$.

Three conclusions follow. First, the proposed GB-SMCF-based GPR provides the strongest overall SE--training trade-off. In particular, the configuration $n_{\textrm{t}}=4$ achieves the highest relative SE while reducing the pilot overhead by $75\%$ and the total training energy by $93.75\%$. This shows that array-geometry-aware Bayesian interpolation can compensate sufficiently for the loss of direct pilot observations such that reduced training outperforms all full-pilot baselines in end-to-end SE. Second, when estimation accuracy is prioritized, the full-training proposed estimator achieves the best NMSE. However, the reduced-training configurations remain highly competitive. In particular, the case $n_{\textrm{t}}=8$ preserves nearly the same relative SE as the full-training proposed configuration, while reducing the pilot overhead by $50\%$ and the total training energy by $75\%$. Thus, $n_{\textrm{t}}=8$ provides the most balanced operating point, whereas $n_{\textrm{t}}=4$ maximizes SE. Third, for the GPR-based estimators, the gain from reduced training is accompanied by a substantial reduction in the dominant exact-GP training cost. As discussed in Section~\ref{Sec:opt_complexity}, the leading term scales as $\mathcal{O}(I_{\textrm{opt}}P^3)$ with $P=N_{\textrm{r}}n_{\textrm{t}}$. Consequently, reducing $n_{\textrm{t}}$ from $16$ to $8$ and $4$ decreases this dominant term by factors of $8$ and $64$, respectively. The proposed configuration with $n_{\textrm{t}}=4$ therefore combines the highest relative SE with a much lower dominant training cost than full-training exact GPR and dense isotropic LMMSE. LS remains substantially cheaper, whereas the OMP and AMP costs additionally depend on the dictionary size $G$ and the iteration counts.

\begin{table}[t]
    \centering
    \caption{Comparison of channel estimators for a $16\times16$ MIMO link.}
    \label{tab:comparison_snr0}
    \scriptsize
    \renewcommand{\arraystretch}{1.15}
    \setlength{\tabcolsep}{1.6pt}
    \resizebox{\columnwidth}{!}{%
    \begin{tabular}{|l|c|c|c|c|c|c|}
        \hline
        \textbf{Estimator}
        & $n_{\textrm{t}}$
        & \shortstack{\textbf{Pil. sav.}\\\textbf{[\%]}}
        & \shortstack{\textbf{En. sav.}\\\textbf{[\%]}}
        & \shortstack{\textbf{Rel. SE}\\\textbf{[\%]}}
        & \shortstack{\textbf{NMSE}\\\textbf{[dB]}}
        & \textbf{Cost} \\
        \hline
        \multicolumn{7}{|c|}{\textit{Proposed GB-SMCF-based GPR}} \\
        \hline
        GB-SMCF GPR & 16 & 0  & 0     & 75.60 & -19.56 & $1.68\times10^7 I_{\textrm{opt}}$ \\
        GB-SMCF GPR & 8  & 50 & 75    & 74.28 & -16.72 & $2.10\times10^6 I_{\textrm{opt}}$ \\
        GB-SMCF GPR & 4  & 75 & 93.75 & 79.99 & -13.78 & $2.62\times10^5 I_{\textrm{opt}}$ \\
        \hline
        \multicolumn{7}{|c|}{\textit{Baseline estimators}} \\
        \hline
        LS              & 16 & 0 & 0 & 45.82 & -10.85 & $2.56\times10^2$ \\
        Isotropic LMMSE & 16 & 0 & 0 & 47.15 & -11.15 & $1.68\times10^7$ \\
        OMP             & 16 & 0 & 0 & 63.70 & -15.52 & $256\,G\,I_{\textrm{OMP}}$ \\
        AMP             & 16 & 0 & 0 & 65.60 & -14.56 & $256\,G\,I_{\textrm{AMP}}$ \\
        \hline
        \multicolumn{7}{|c|}{\textit{Learning-based GPR with standard kernels}} \\
        \hline
        SqExp-GPR   & 16 & 0  & 0  & 66.52 & -13.25 & $1.68\times10^7 I_{\textrm{opt}}$ \\
        SqExp-GPR   & 8  & 50 & 75 & 70.23 & -11.13 & $2.10\times10^6 I_{\textrm{opt}}$ \\
        Mat\'ern-GPR & 16 & 0  & 0  & 67.09 & -15.28 & $1.68\times10^7 I_{\textrm{opt}}$ \\
        Mat\'ern-GPR & 8  & 50 & 75 & 72.37 & -12.39 & $2.10\times10^6 I_{\textrm{opt}}$ \\
        \hline
    \end{tabular}%
    }
\end{table}

%%%%%%%%%%%%%%%%%%%%%%%%%%%%%%%%%%%%%%%%%%%%%%%%%%%%%%%%%%%%%%%%%%%%%%%%%%%%%%%%%%%%%%%%%%
%%%%%%%%%%%%%%%%%%%%%%%%%%%%%%%%%%%%%%%%%%%%%%%%%%%%%%%%%%%%%%%%%%%%%%%%%%%%%%%%%%%%%%%%%%
\vspace{-2mm}
\section{Conclusion and Future Directions}
\label{sec:conclusion}
This paper addressed pilot-limited channel estimation as a highly underdetermined Bayesian inverse problem and proposed a GB-SMCF-based GPR estimator to solve it. The wireless channel is modeled as a proper complex Gaussian random field over the joint transmit--receive antenna-index lattice, with the proposed GB-SMCF prior covariance kernel, which captures smooth spatial decay, oscillatory correlation patterns, and anisotropic behavior induced by array geometry and multipath propagation. This learned kernel is embedded into a GPR framework to reconstruct the full CSI from sparse noisy pilot observations. The resulting posterior mean smooths the observed channel entries and predicts the unobserved entries, while the posterior covariance quantifies the reconstruction uncertainty.

The proposed GB-SMCF-based GPR estimator circumvents multiple restrictive assumptions of conventional methods: unlike CS-based estimators, it requires no fixed dictionary, grid-aligned angular sparsity, or strict low-rank structure; and unlike covariance-based MMSE estimators, it does not require the channel covariance to be known. Instead, it learns a physically interpretable array-domain covariance model directly from the current pilot observations within each coherence block. Numerical results under the Saleh--Valenzuela and 3GPP TR~38.901 CDL-A channel models show that it reduces pilot overhead by up to $75\%$ and total training energy by up to $93.75\%$, while achieving lower NMSE and higher SE than the benchmark techniques.

Future work will leverage the posterior predictive uncertainty for adaptive pilot allocation, reliability-aware link adaptation, and uncertainty-aware beamforming. Another important direction is to extend the prior to handle non-stationary effects in larger arrays, for instance via spatially varying hyperparameters or local kernel mixtures.

\appendices
\vspace{-2mm}
\section{Proof of Lemma~\ref{lem:PSD}}
\label{app:psd-proof}

We prove: i) lattice stationarity of $k_{\textrm{GB}}$, ii) Hermitian positive semidefiniteness of $k_{\textrm{GB}}$, and iii) positive semidefiniteness of its real-augmented covariance.

\paragraph{Lattice stationarity}
By~\eqref{eq:K-rx} and~\eqref{eq:K-tx}, $k_{\textrm{r}}(i,i')$ depends only on $\Delta\mathbf{r}=\mathbf{r}_i-\mathbf{r}_{i'}=(\Delta r_y,\Delta r_z)$, and $k_{\textrm{t}}(j,j')$ depends only on $\Delta\mathbf{t}=\mathbf{t}_j-\mathbf{t}_{j'}=(\Delta t_y,\Delta t_z)$. Hence, by~\eqref{eq:GB-SM-function}, $k_{\textrm{GB}}\big((i,j),(i',j')\big) = \alpha\,k_{\textrm{r}}(\Delta\mathbf{r})\,k_{\textrm{t}}(\Delta\mathbf{t})$, so $k_{\textrm{GB}}$ depends only on the joint lattice difference $(\Delta\mathbf{r},\Delta\mathbf{t})$.

\paragraph{Hermitian PSD of the side kernels}
Fix $s\in\{\textrm{r},\textrm{t}\}$. Each component of the side kernel in~\eqref{eq:K-rx} or~\eqref{eq:K-tx} has the form $\exp\!\Big(-2\pi^2\big[v_{q,y}^{(s)}(\Delta u_y)^2+v_{q,z}^{(s)}(\Delta u_z)^2\big]\Big) \exp\!\Big(\mathrm{j}\,2\pi\big[\mu_{q,y}^{(s)}\Delta u_y+\mu_{q,z}^{(s)}\Delta u_z\big]\Big)$, which is the inverse Fourier transform of a nonnegative Gaussian spectral density. Therefore, by Bochner's theorem, each component is a Hermitian PSD kernel on $\R^2$. Restricting a PSD kernel to the integer lattice preserves positive semidefiniteness, and nonnegative weighted sums preserve positive semidefiniteness. Hence the receive-side and transmit-side kernels are Hermitian PSD on their respective lattices.

\paragraph{Hermitian PSD of the GB-SMCF}
Fix a finite ordered set $\mathcal{X}=\{(i_a,j_a)\}_{a=1}^P\subset\mathcal{G}$, and define $[\mathbf{K}_{\textrm{r}}]_{ab}=k_{\textrm{r}}(i_a,i_b)$, \qquad $[\mathbf{K}_{\textrm{t}}]_{ab}=k_{\textrm{t}}(j_a,j_b)$. From the previous paragraph, $\mathbf{K}_{\textrm{r}}\succeq\mathbf{0}$ and $\mathbf{K}_{\textrm{t}}\succeq\mathbf{0}$, and both are Hermitian. By~\eqref{eq:GB-SM-function}, $\mathbf{K}_{\textrm{GB}}(\mathcal{X},\mathcal{X}) = \alpha(\mathbf{K}_{\textrm{r}}\circ\mathbf{K}_{\textrm{t}})$, where $\circ$ denotes the Hadamard product. Since $\alpha>0$ and the Hadamard product of Hermitian PSD matrices is Hermitian PSD by the Schur product theorem, $\mathbf{K}_{\textrm{GB}}(\mathcal{X},\mathcal{X})\succeq\mathbf{0}$. Hence $k_{\textrm{GB}}$ is a Hermitian PSD kernel on $\mathcal{G}\times\mathcal{G}$.

\paragraph{PSD of the real-augmented covariance}
Let $\mathbf{K}=\mathbf{K}_{\textrm{GB}}(\mathcal{X},\mathcal{X})$ and write $\mathbf{K}=\mathbf{A}+\mathrm{j}\,\mathbf{B}$, where $\mathbf{A}=\Re\{\mathbf{K}\}$ and $\mathbf{B}=\Im\{\mathbf{K}\}$. Since $\mathbf{K}$ is Hermitian, $\mathbf{A}$ is symmetric and $\mathbf{B}$ is skew-symmetric. For any $\mathbf{u},\mathbf{v}\in\R^P$, define $\mathbf{x}=[\mathbf{u}^{\mathsf{T}}\ \mathbf{v}^{\mathsf{T}}]^{\mathsf{T}}$ and $\mathbf{c}=\mathbf{u}+\mathrm{j}\,\mathbf{v}\in\C^{P}$. Using~\eqref{eq:T-operator}, $\mathbf{x}^{\mathsf{T}}\mathcal{T}(\mathbf{K})\mathbf{x} = \frac{1}{2} \begin{bmatrix} \mathbf{u}^{\mathsf{T}} & \mathbf{v}^{\mathsf{T}} \end{bmatrix} \begin{bmatrix} \mathbf{A} & -\mathbf{B}\\ \mathbf{B} & \mathbf{A} \end{bmatrix} \begin{bmatrix} \mathbf{u}\\ \mathbf{v} \end{bmatrix} = \frac{1}{2}\mathbf{c}^{\Hsf}\mathbf{K}\mathbf{c} \ge 0$, because $\mathbf{K}\succeq\mathbf{0}$. Therefore, $\mathcal{T}\!\big(\mathbf{K}_{\textrm{GB}}(\mathcal{X},\mathcal{X})\big)\succeq\mathbf{0}$. Finally, $\boldsymbol{\Sigma}_{\textrm{n}} = \frac{\sigma_{\textrm{obs}}^2}{2}\I_{2P}\succeq\mathbf{0}$, so $\mathcal{T}\!\big(\mathbf{K}_{\textrm{GB}}(\mathcal{X},\mathcal{X})\big)+\boldsymbol{\Sigma}_{\textrm{n}}\succeq\mathbf{0}$. This proves Lemma~\ref{lem:PSD}.

\vspace{-2mm}
\section{Closed-Form GB-SMCF Derivatives}
\label{app:kernel-grads}

This appendix gives the derivatives required in~\eqref{eq:GP-grad-master} for the GB-SMCF in~\eqref{eq:K-rx}--\eqref{eq:GB-SM-function}. Since optimization is performed over the optimization variables $\theta$, all derivatives below are given with respect to those variables.

\paragraph{Side-kernel component notation}
For $s\in\{\textrm{r},\textrm{t}\}$, let $\Delta\mathbf{u}=(\Delta u_y,\Delta u_z)$ denote the corresponding lattice difference, i.e., $\Delta\mathbf{u}=\Delta\mathbf{r}$ for $s=\textrm{r}$ and $\Delta\mathbf{u}=\Delta\mathbf{t}$ for $s=\textrm{t}$. For a component index $q\in\{1,\dots,Q_s\}$, define $E_q^{(s)} = \exp\!\Big(-2\pi^2\big[v_{q,y}^{(s)}(\Delta u_y)^2+v_{q,z}^{(s)}(\Delta u_z)^2\big]\Big)$, $\phi_q^{(s)} = 2\pi\big[\mu_{q,y}^{(s)}\Delta u_y+\mu_{q,z}^{(s)}\Delta u_z\big]$, $G_q^{(s)} = E_q^{(s)}\exp\!\big(\mathrm{j}\,\phi_q^{(s)}\big)$, \quad $C_q^{(s)} = \pi_q^{(s)}G_q^{(s)}$. Then $k_s(\cdot,\cdot)=\sum_{q=1}^{Q_s}C_q^{(s)}$.

\paragraph{Per-component derivatives with respect to optimization variables}
For each side $s\in\{\textrm{r},\textrm{t}\}$ and each $q=1,\dots,Q_s-1$, the derivative with respect to the softmax reparameterization $\theta_{\pi,q}^{(s)}$ is
\begin{equation}
\label{eq:deriv-logits}
\frac{\partial k_s}{\partial \theta_{\pi,q}^{(s)}}
=
\pi_q^{(s)}\big(G_q^{(s)}-k_s\big)
=
C_q^{(s)}-\pi_q^{(s)}k_s.
\end{equation}
For each $q=1,\dots,Q_s$ and $\xi\in\{y,z\}$, the derivatives with respect to the spatial-frequency and spectral-variance optimization variables are
\begin{equation}
\label{eq:deriv-mu-theta}
\frac{\partial k_s}{\partial \theta_{\mu,q,\xi}^{(s)}}
=
\mathrm{j}\,\pi\bigl(1-4(\mu_{q,\xi}^{(s)})^2\bigr)\,\Delta u_\xi\,C_q^{(s)},
\end{equation}
\begin{equation}
\label{eq:deriv-v-theta}
\frac{\partial k_s}{\partial \theta_{v,q,\xi}^{(s)}}
=
-2\pi^2\,v_{q,\xi}^{(s)}(\Delta u_\xi)^2\,C_q^{(s)}.
\end{equation}
Equations~\eqref{eq:deriv-mu-theta} and~\eqref{eq:deriv-v-theta} follow from $\frac{\partial \mu_{q,\xi}^{(s)}}{\partial \theta_{\mu,q,\xi}^{(s)}} = \frac{1}{2}\Big(1-\tanh^2(\theta_{\mu,q,\xi}^{(s)})\Big) = \frac{1}{2}\big(1-4(\mu_{q,\xi}^{(s)})^2\big)$ and $\frac{\partial v_{q,\xi}^{(s)}}{\partial \theta_{v,q,\xi}^{(s)}} = v_{q,\xi}^{(s)}$.

\paragraph{Base-kernel derivatives}
For $x=(i,j)$ and $x'=(i',j')$, the GB-SMCF satisfies $k_{\textrm{GB}}(x,x')=\alpha\,k_{\textrm{r}}(i,i')\,k_{\textrm{t}}(j,j')$. Hence,
\begin{equation}
\label{eq:app-dK-product}
\frac{\partial k_{\textrm{GB}}}{\partial \theta_\alpha}
=
k_{\textrm{GB}},
\;
\frac{\partial k_{\textrm{GB}}}{\partial \theta_{\textrm{r}}}
=
\alpha\,\frac{\partial k_{\textrm{r}}}{\partial \theta_{\textrm{r}}}\,k_{\textrm{t}},
\;
\frac{\partial k_{\textrm{GB}}}{\partial \theta_{\textrm{t}}}
=
\alpha\,k_{\textrm{r}}\,\frac{\partial k_{\textrm{t}}}{\partial \theta_{\textrm{t}}},
\end{equation}
where $\theta_{\textrm{r}}$ and $\theta_{\textrm{t}}$ denote arbitrary receive-side and transmit-side optimization variables, respectively.

\paragraph{Matrix derivatives}
For a general training set $\mathcal{X}=\{(i,a_\ell)\}_{p=1}^P$, let $[\mathbf{K}_{\textrm{r}}]_{ab}=k_{\textrm{r}}(i_a,i_b)$, \quad $[\mathbf{K}_{\textrm{t}}]_{ab}=k_{\textrm{t}}(j_a,j_b)$, \quad $\mathbf{K}_{\textrm{GB}}=\alpha(\mathbf{K}_{\textrm{r}}\circ\mathbf{K}_{\textrm{t}})$. Then
\begin{equation}
\resizebox{\columnwidth}{!}{$
\label{eq:matrix-derivs}
\frac{\partial \mathbf{K}_{\textrm{GB}}}{\partial \theta_\alpha}
=
\mathbf{K}_{\textrm{GB}},
\;
\frac{\partial \mathbf{K}_{\textrm{GB}}}{\partial \theta_{\textrm{r}}}
=
\alpha\Big(\frac{\partial \mathbf{K}_{\textrm{r}}}{\partial \theta_{\textrm{r}}}\circ\mathbf{K}_{\textrm{t}}\Big),
\;
\frac{\partial \mathbf{K}_{\textrm{GB}}}{\partial \theta_{\textrm{t}}}
=
\alpha\Big(\mathbf{K}_{\textrm{r}}\circ\frac{\partial \mathbf{K}_{\textrm{t}}}{\partial \theta_{\textrm{t}}}\Big).$}
\end{equation}

\paragraph{Real-augmented covariance derivatives}
With $\widetilde{\mathbf{K}}_{\mathcal{X}\mathcal{X}} = \mathcal{T}(\mathbf{K}_{\textrm{GB}})$, \quad $\mathbf{C}_\theta = \widetilde{\mathbf{K}}_{\mathcal{X}\mathcal{X}}+\boldsymbol{\Sigma}_{\textrm{n}}$, linearity of $\mathcal{T}(\cdot)$ yields
\begin{equation}
\label{eq:aug-derivs}
\frac{\partial \widetilde{\mathbf{K}}_{\mathcal{X}\mathcal{X}}}{\partial \theta_m}
=
\mathcal{T}\!\left(
\frac{\partial \mathbf{K}_{\textrm{GB}}}{\partial \theta_m}
\right),
\quad
\frac{\partial \mathbf{C}_\theta}{\partial \theta_m}
=
\mathcal{T}\!\left(
\frac{\partial \mathbf{K}_{\textrm{GB}}}{\partial \theta_m}
\right),
\end{equation}
because $\boldsymbol{\Sigma}_{\textrm{n}}$ is fixed in the present formulation. Substituting~\eqref{eq:aug-derivs} into~\eqref{eq:GP-grad-master} gives the exact gradients used in Algorithm~\ref{alg:cov_design}.

\bibliographystyle{IEEEtran_renamed.bst}
\bibliography{bibliography.bib}

\begin{IEEEbiography}
[{\includegraphics[width=1in,height=1.25in,clip]{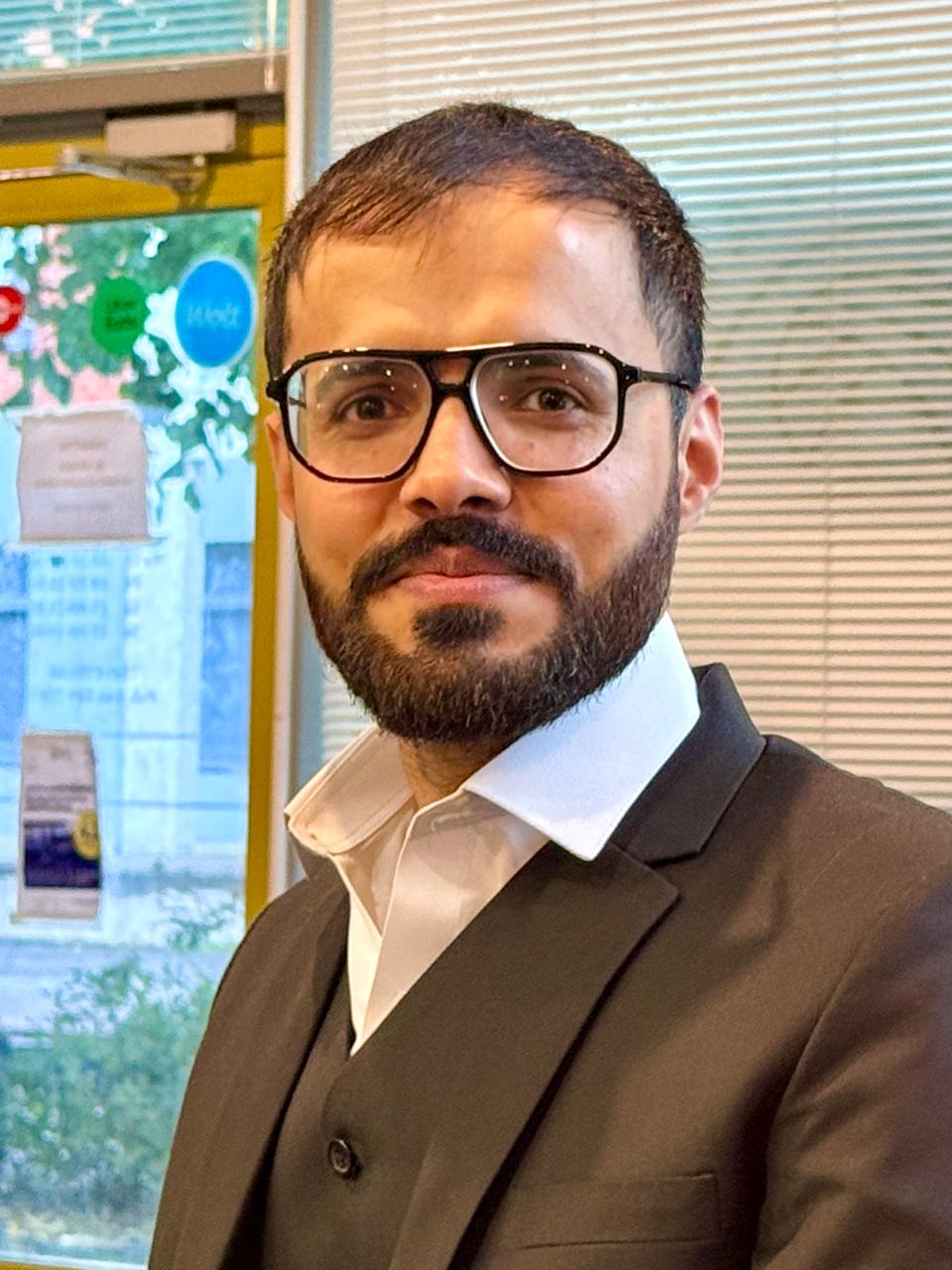}}]
{Syed Luqman Shah} (Graduate Student Member, IEEE) received the B.Sc. degree in electrical engineering from the UET, Peshawar, Pakistan, in 2021, with the University Gold Medal, and the M.Sc. degree in electrical engineering with distinction from the GIKI of Engineering Sciences and Technology, Pakistan, in 2023. From 2023 to 2024, he was a Research Assistant with FAST--NUCES, Islamabad, Pakistan. He is currently pursuing the doctoral degree with the Centre for Wireless Communications, University of Oulu, Finland. His research interests include wireless propagation modeling, wireless digital twins, channel estimation, and hyper-reliable low-latency communications.
\end{IEEEbiography}

\begin{IEEEbiography}
  [{\includegraphics[width=1in,height=1.25in,clip]{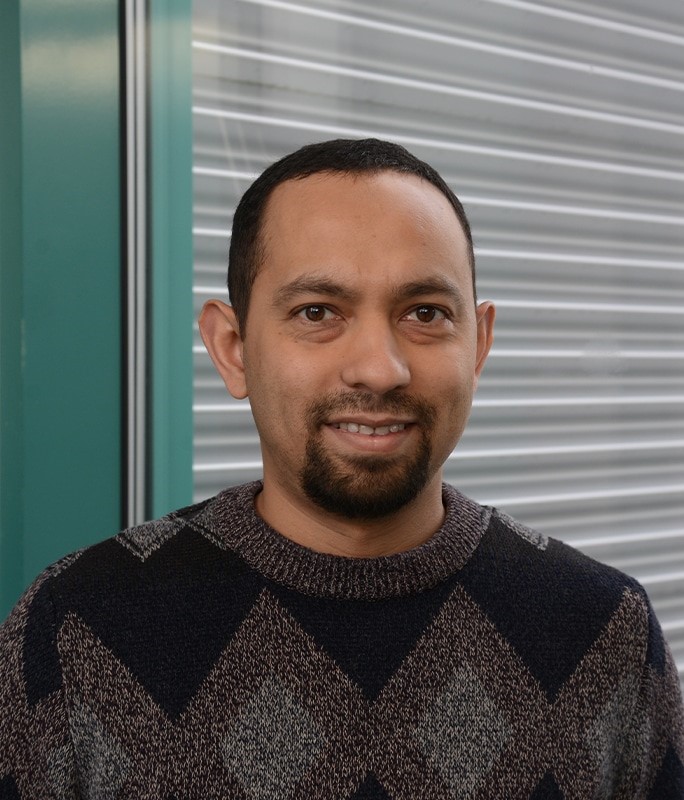}}]
  {Nurul Huda Mahmood} (Member, IEEE)  was born in Bangladesh. He received the Ph.D. degree from NTNU, Norway, in 2013. He is currently a Senior Research Fellow with CWC, University of Oulu, Finland, where he coordinates the wireless connectivity research area in the Finnish 6G Flagship program and leads the critical MTC research group. Before joining the University of Oulu in 2018, he was an Associate Professor with the Department of Electronics Systems, Aalborg University, Denmark. He has contributed to various international research projects including Hexa-X-II - EU’s flagship 6G research project. He has co-authored over 100 peer-reviewed publications. His current research interests include resilient communications for wireless networks.
\end{IEEEbiography}%

\end{document}

%% file: Figures/contour_init_sensitivity.tex
% ---------------------------------------------------------------------------
% Consistent colors used 
% ---------------------------------------------------------------------------
\definecolor{trajOne}{RGB}{31,119,180}     % Start 1
\definecolor{trajTwo}{RGB}{214,39,40}      % Start 2
\definecolor{trajThree}{RGB}{227,119,194}  % Start 3
\definecolor{trajFour}{RGB}{23,190,207}    % Start 4
\definecolor{trajFive}{RGB}{44,160,44}      % Start 5
\definecolor{trajSix}{RGB}{140,86,75}       % Start 6
\definecolor{startDot}{RGB}{220,50,47}      % common start-point marker
\definecolor{endCross}{RGB}{255,127,14}     % common end-point marker

% ---------------------------------------------------------------------------
% ---------------------------------------------------------------------------
\pgfplotsset{
    multistart1/.style={draw=trajOne,   line width=1.05pt},
    multistart2/.style={draw=trajTwo,   line width=1.05pt},
    multistart3/.style={draw=trajThree, line width=1.05pt},
    multistart4/.style={draw=trajFour,  line width=1.05pt},
    multistart5/.style={draw=trajFive,  line width=1.05pt},
    multistart6/.style={draw=trajSix,   line width=1.05pt}
}

\begin{tikzpicture}

% ---------------------------------------------------------------------------
% LEFT SUBFIGURE: local objective landscape and projected optimizer paths
% ---------------------------------------------------------------------------
\begin{axis}[
    name=leftpanel,
    width=3.0cm,
    height=6.8cm,
    at={(0cm,0cm)},
    scale only axis,
    view={0}{90},
    xmin=-0.30, xmax=0.32,
    ymin=-0.33, ymax=0.30,
    xlabel={$\mu^{(\mathrm{t})}_{2,y}$},
    ylabel={$\mu^{(\mathrm{t})}_{2,z}$},
    xlabel near ticks,
    ylabel near ticks,
    xlabel style={font=\footnotesize, yshift=1.0mm},
    ylabel style={font=\footnotesize, xshift=1.0mm, yshift=-5mm},
    ticklabel style={font=\scriptsize},
    xtick={-0.2,0.0,0.2},
    ytick={-0.3,-0.2,-0.1,0,0.1,0.2,0.3},
    grid=major,
    major grid style={draw=white!28, line width=0.35pt},
    axis line style={black},
    tick style={black},
    colormap/viridis,
    point meta min=1.12483,
    point meta max=3.13583,
    colorbar,
    colorbar style={
        width=0.30cm,
        xshift=-2.6mm,
        ytick={1.12483,2.13037,3.09583},
        yticklabels={$1.33\times 10^{1}$,$1.35\times 10^{2}$,$1.37\times 10^{3}$},
        yticklabel pos=right,
        yticklabel style={font=\scriptsize, rotate=90, anchor=center, xshift=-2.7mm, yshift=-0.9mm},
        ylabel={Marginal-likelihood gap $\Delta\mathcal{L}(\theta)$},
        ylabel style={font=\footnotesize, yshift=2.0mm}
    },
    legend style={
        at={(0.50,0.985)},
        anchor=north,
        font=\scriptsize,
        fill=white,
        fill opacity=0.88,
        draw=black!20,
        text opacity=1,
        inner sep=1.5pt,
        legend columns=2,
        /tikz/every even column/.append style={column sep=1.6mm}
    },
    legend cell align=left,
    unbounded coords=jump
]

% Heatmap of the marginal-likelihood gap.
\addplot3[
    surf,
    shader=interp,
    draw=none,
    mesh/rows=61,
    point meta=explicit,
    forget plot
] table [x=x, y=y, z=z, meta expr={ln(\thisrow{z})/ln(10)}, col sep=comma]
{Figures/files_txt/contour_init_sensitivity_heatmap.txt};

\foreach \a/\b in {0.055/0.070,0.075/0.105,0.095/0.138,0.120/0.170,0.150/0.205,0.185/0.240,0.225/0.275,0.270/0.315,0.320/0.355,0.375/0.395,0.440/0.440} {
    \addplot[
        black,
        line width=0.8pt,
        samples=220,
        domain=0:360,
        smooth,
        forget plot
    ] ({-0.20 + 0.015*sin(x) + \a*cos(x)}, {0.00 + \b*sin(x)});
}

\addplot[
    only marks,
    mark=star,
    mark size=8.5pt,
    mark options={solid, fill=red, draw=red},
    forget plot
] coordinates {(-0.205,0.0)};
\addlegendimage{only marks, mark=star, mark size=3.8pt, mark options={solid, fill=red, draw=red}}
\addlegendentry{Best fit}

% Six projected multiStart trajectories.
\addplot[multistart1] table [x=start1_x, y=start1_y, col sep=comma] {Figures/files_txt/contour_init_sensitivity_paths.txt};
\addlegendentry{Start 1}
\addplot[multistart2] table [x=start2_x, y=start2_y, col sep=comma] {Figures/files_txt/contour_init_sensitivity_paths.txt};
\addlegendentry{Start 2}
\addplot[multistart3] table [x=start3_x, y=start3_y, col sep=comma] {Figures/files_txt/contour_init_sensitivity_paths.txt};
\addlegendentry{Start 3}
\addplot[multistart4] table [x=start4_x, y=start4_y, col sep=comma] {Figures/files_txt/contour_init_sensitivity_paths.txt};
\addlegendentry{Start 4}
\addplot[multistart5] table [x=start5_x, y=start5_y, col sep=comma] {Figures/files_txt/contour_init_sensitivity_paths.txt};
\addlegendentry{Start 5}
\addplot[multistart6] table [x=start6_x, y=start6_y, col sep=comma] {Figures/files_txt/contour_init_sensitivity_paths.txt};
\addlegendentry{Start 6}

% Start points S_i: common red filled dots.
\addplot[
    only marks,
    mark=*,
    mark size=2.6pt,
    mark options={solid, fill=startDot, draw=startDot},
    forget plot
] coordinates {
    (0.09,0.31) (0.26,0.05) (0.30,-0.18) (0.22,0.205) (0.25,-0.30) (-0.05,0.30)
};

\addplot[
    only marks,
    mark=x,
    mark size=3.1pt,
    line width=0.9pt,
    endCross,
    forget plot
] coordinates {
    (-0.05,-0.01) (-0.08,-0.07) (-0.03,0.01) (-0.05,0.01) (-0.10,-0.08) (-0.12,0.03)
};

% Labels for start points.
\node[font=\tiny, fill=white, inner sep=0.6pt, anchor=west] at (axis cs:0.098,0.315) {$S_1$};
\node[font=\tiny, fill=white, inner sep=0.6pt, anchor=west] at (axis cs:0.252,0.065) {$S_2$};
\node[font=\tiny, fill=white, inner sep=0.6pt, anchor=west] at (axis cs:0.215,-0.178) {$S_3$};
\node[font=\tiny, fill=white, inner sep=0.6pt, anchor=west] at (axis cs:0.244,0.210) {$S_4$};
\node[font=\tiny, fill=white, inner sep=0.6pt, anchor=west] at (axis cs:0.238,-0.282) {$S_5$};
\node[font=\tiny, fill=white, inner sep=0.6pt, anchor=west] at (axis cs:-0.040,0.289) {$S_6$};

% Labels for end points.
\node[font=\tiny, fill=white, inner sep=0.6pt, anchor=east] at (axis cs:-0.058,-0.016) {$E_1$};
\node[font=\tiny, fill=white, inner sep=0.6pt, anchor=east] at (axis cs:-0.087,-0.065) {$E_2$};
\node[font=\tiny, fill=white, inner sep=0.6pt, anchor=east] at (axis cs:-0.027,0.018) {$E_3$};
\node[font=\tiny, fill=white, inner sep=0.6pt, anchor=east] at (axis cs:0.058,0.021) {$E_4$};
\node[font=\tiny, fill=white, inner sep=0.6pt, anchor=east] at (axis cs:-0.108,-0.085) {$E_5$};
\node[font=\tiny, fill=white, inner sep=0.6pt, anchor=east] at (axis cs:-0.128,0.036) {$E_6$};
\end{axis}

% ---------------------------------------------------------------------------
% RIGHT SUBFIGURE: convergence-gap curves
% ---------------------------------------------------------------------------
\begin{axis}[
    name=rightpanel,
    width=3.0cm,
    height=6.8cm,
    at={(4.9cm,0cm)},
    scale only axis,
    xmin=-0.5, xmax=40.5,
    ymin=2e-2, ymax=1e7,
    ymode=log,
    xlabel={Optimizer iteration},
    ylabel={Convergence gap $\mathcal{L}(\hat{\theta})-\mathcal{L}(\theta^{(k)})$},
    xlabel near ticks,
    ylabel near ticks,
    xlabel style={font=\footnotesize, yshift=1.0mm},
    ylabel style={font=\footnotesize, xshift=1.0mm, yshift=-4.5mm},
    ticklabel style={font=\scriptsize},
    xtick={0,10,20,30,40},
    ytickten={-1,0,1,2,3,4,5,6,7},
    grid=major,
    major grid style={draw=black!15, line width=0.35pt},
    axis line style={black},
    tick style={black},
    legend style={
        at={(0.50,0.985)},
        anchor=north,
        font=\scriptsize,
        fill=white,
        fill opacity=0.90,
        draw=black!20,
        text opacity=1,
        inner sep=1.5pt,
        legend columns=2,
        /tikz/every even column/.append style={column sep=1.5mm}
    },
    legend cell align=left
]

\addplot[multistart1] table [x=iter, y=start1, col sep=comma] {Figures/files_txt/contour_init_sensitivity_convergence.txt};
\addlegendentry{Start 1}
\addplot[multistart2] table [x=iter, y=start2, col sep=comma] {Figures/files_txt/contour_init_sensitivity_convergence.txt};
\addlegendentry{Start 2}
\addplot[multistart3] table [x=iter, y=start3, col sep=comma] {Figures/files_txt/contour_init_sensitivity_convergence.txt};
\addlegendentry{Start 3}
\addplot[multistart4] table [x=iter, y=start4, col sep=comma] {Figures/files_txt/contour_init_sensitivity_convergence.txt};
\addlegendentry{Start 4}
\addplot[multistart5] table [x=iter, y=start5, col sep=comma] {Figures/files_txt/contour_init_sensitivity_convergence.txt};
\addlegendentry{Start 5}
\addplot[multistart6] table [x=iter, y=start6, col sep=comma] {Figures/files_txt/contour_init_sensitivity_convergence.txt};
\addlegendentry{Start 6}
\end{axis}
\end{tikzpicture}

%% file: Figures/nmse_plot_SV_16.tex
% ============================================================================
\definecolor{plotRed}{RGB}{214,39,40}       % GB-SMCF based GPR, n_t = 8
\definecolor{plotGreen}{RGB}{0,158,115}     % GB-SMCF based GPR, n_t = 4
\definecolor{plotOrange}{RGB}{213,94,0}     % LMMSE-isotropic
\definecolor{plotCyan}{RGB}{23,190,207}     % OMP
\definecolor{plotGold}{RGB}{230,159,0}      % AMP
% ============================================================================
\pgfdeclareplotmark{nmseOpenPlus}{%
    \pgfpathmoveto{\pgfpoint{-0.62pt}{ 1.85pt}}%
    \pgfpathlineto{\pgfpoint{ 0.62pt}{ 1.85pt}}%
    \pgfpathlineto{\pgfpoint{ 0.62pt}{ 0.62pt}}%
    \pgfpathlineto{\pgfpoint{ 1.85pt}{ 0.62pt}}%
    \pgfpathlineto{\pgfpoint{ 1.85pt}{-0.62pt}}%
    \pgfpathlineto{\pgfpoint{ 0.62pt}{-0.62pt}}%
    \pgfpathlineto{\pgfpoint{ 0.62pt}{-1.85pt}}%
    \pgfpathlineto{\pgfpoint{-0.62pt}{-1.85pt}}%
    \pgfpathlineto{\pgfpoint{-0.62pt}{-0.62pt}}%
    \pgfpathlineto{\pgfpoint{-1.85pt}{-0.62pt}}%
    \pgfpathlineto{\pgfpoint{-1.85pt}{ 0.62pt}}%
    \pgfpathlineto{\pgfpoint{-0.62pt}{ 0.62pt}}%
    \pgfpathclose%
    \pgfusepath{fill,stroke}%
}

\pgfdeclareplotmark{nmseOpenX}{%
    \pgfpathmoveto{\pgfpoint{-0.93pt}{ 1.85pt}}%
    \pgfpathlineto{\pgfpoint{ 0.00pt}{ 0.93pt}}%
    \pgfpathlineto{\pgfpoint{ 0.93pt}{ 1.85pt}}%
    \pgfpathlineto{\pgfpoint{ 1.85pt}{ 0.93pt}}%
    \pgfpathlineto{\pgfpoint{ 0.93pt}{ 0.00pt}}%
    \pgfpathlineto{\pgfpoint{ 1.85pt}{-0.93pt}}%
    \pgfpathlineto{\pgfpoint{ 0.93pt}{-1.85pt}}%
    \pgfpathlineto{\pgfpoint{ 0.00pt}{-0.93pt}}%
    \pgfpathlineto{\pgfpoint{-0.93pt}{-1.85pt}}%
    \pgfpathlineto{\pgfpoint{-1.85pt}{-0.93pt}}%
    \pgfpathlineto{\pgfpoint{-0.93pt}{ 0.00pt}}%
    \pgfpathlineto{\pgfpoint{-1.85pt}{ 0.93pt}}%
    \pgfpathclose%
    \pgfusepath{fill,stroke}%
}

\begin{tikzpicture}

\begin{axis}[
    % ========================================================================
    % Figure size
    % ========================================================================
    width=8.5cm, height=7.5cm,
    %
    % ========================================================================
    % Axis limits
    % ========================================================================
    xmin=-16.25, xmax=11.25,
    ymin=-26.574, ymax=4.592,
    %
    % ========================================================================
    % Axis labels
    % ========================================================================
    xlabel={SNR [dB]},
    ylabel={NMSE (dB)},
    xlabel near ticks,
    ylabel near ticks,
    label style={font=\footnotesize},
    %
    % ========================================================================
    % Major ticks
    % ========================================================================
    xtick={-15,-10,...,10},
    ytick={-25,-20,...,0},
    ticklabel style={font=\footnotesize},
    %
    % ========================================================================
    % Legend
    % ========================================================================
    legend style={
        at={(0.98,0.98)},
        anchor=north east,
        font=\scriptsize,
        inner sep=1pt,
        fill opacity=0.85,
        draw=black!20,
        draw opacity=1,
        text opacity=1
    },
    legend cell align=left,
    %
    % ========================================================================
    % Grid
    % ========================================================================
    grid=major,
    major grid style={
        draw=black!19,
        line width=0.3pt,
        dash pattern=on 0.3pt off 0.5pt
    },
    mark options={solid}
]

% ============================================================================
% Curve 1: GB-SMCF based GPR, n_t = 8
% ============================================================================
\addplot[
    line width=1pt,
    plotRed,
    solid,
    mark=square,
    mark size=1.85pt,
    mark options={solid, fill=white, line width=0.6pt}
]
table [x=SNR_dB, y=GB_SMCF_GPR_nt8, col sep=comma]
{Figures/files_txt/nmse_plot_SV_16.txt};
\addlegendentry{GB-SMCF based GPR ($n_{\textrm{t}}=8$)}

% ============================================================================
% Curve 2: GB-SMCF based GPR, n_t = 4
% ============================================================================
\addplot[
    line width=1pt,
    plotGreen,
    dash pattern=on 6.6pt off 1.65pt on 1.05pt off 1.65pt,
    mark=triangle,
    mark size=2.1pt,
    mark options={solid, fill=white, line width=0.6pt}
]
table [x=SNR_dB, y=GB_SMCF_GPR_nt4, col sep=comma]
{Figures/files_txt/nmse_plot_SV_16.txt};
\addlegendentry{GB-SMCF based GPR ($n_{\textrm{t}}=4$)}

% ============================================================================
% Curve 3: AMP, n_t = 16
% ============================================================================
\addplot[
    line width=1pt,
    plotGold,
    dash pattern=on 3.10pt off 1.05pt on 1.05pt off 1.05pt,
    mark=nmseOpenX,
    mark options={solid, fill=white, line width=0.6pt}
]
table [x=SNR_dB, y=AMP_nt16, col sep=comma]
{Figures/files_txt/nmse_plot_SV_16.txt};
\addlegendentry{AMP ($n_{\textrm{t}}=16$)}

% ============================================================================
% Curve 4: OMP, n_t = 16
% ============================================================================
\addplot[
    line width=1pt,
    plotCyan,
    dash pattern=on 1.05pt off 1.70pt,
    mark=triangle,
    mark size=2.1pt,
    mark options={solid, fill=white, line width=0.6pt, rotate=180}
]
table [x=SNR_dB, y=OMP_nt16, col sep=comma]
{Figures/files_txt/nmse_plot_SV_16.txt};
\addlegendentry{OMP ($n_{\textrm{t}}=16$)}

% ============================================================================
% Curve 5: LMMSE-isotropic, n_t = 16
% ============================================================================
\addplot[
    line width=1pt,
    plotOrange,
    dash pattern=on 6.6pt off 1.65pt on 1.05pt off 1.65pt,
    mark=nmseOpenPlus,
    mark options={solid, fill=white, line width=0.6pt}
]
table [x=SNR_dB, y=LMMSE_isotropic_nt16, col sep=comma]
{Figures/files_txt/nmse_plot_SV_16.txt};
\addlegendentry{LMMSE-isotropic ($n_{\textrm{t}}=16$)}

% ============================================================================
% Curve 6: LS, n_t = 16
% ============================================================================
\addplot[
    line width=1pt,
    black,
    dash pattern=on 3.85pt off 1.65pt,
    mark=diamond,
    mark size=2.55pt,
    mark options={solid, fill=white, line width=0.6pt}
]
table [x=SNR_dB, y=LS_nt16, col sep=comma]
{Figures/files_txt/nmse_plot_SV_16.txt};
\addlegendentry{LS ($n_{\textrm{t}}=16$)}

\end{axis}
\end{tikzpicture}

%% file: Figures/nmse_plot_SV_GB_SM_Mat_SE_16.tex
% ============================================================================

% ============================================================================
% Exact RGB colors
% ============================================================================
\definecolor{plotBlue}{RGB}{0,114,178}      % GB-SMCF based GPR, n_t = 16
\definecolor{plotRed}{RGB}{214,39,40}       % GB-SMCF based GPR, n_t = 8
\definecolor{plotGreen}{RGB}{0,158,115}     % GB-SMCF based GPR, n_t = 4
\definecolor{plotPurple}{RGB}{127,60,141}   % Matern based GPR, n_t = 16
\definecolor{plotBrown}{RGB}{140,86,75}     % SqExp based GPR, n_t = 16

% ============================================================================
% ============================================================================
\pgfdeclareplotmark{nmseOpenHexagon}{%
    \pgfpathmoveto{\pgfpoint{ 0.00pt}{ 1.85pt}}%
    \pgfpathlineto{\pgfpoint{ 1.60pt}{ 0.925pt}}%
    \pgfpathlineto{\pgfpoint{ 1.60pt}{-0.925pt}}%
    \pgfpathlineto{\pgfpoint{ 0.00pt}{-1.85pt}}%
    \pgfpathlineto{\pgfpoint{-1.60pt}{-0.925pt}}%
    \pgfpathlineto{\pgfpoint{-1.60pt}{ 0.925pt}}%
    \pgfpathclose%
    \pgfusepath{fill,stroke}%
}

\begin{tikzpicture}

\begin{axis}[
    % ========================================================================
    % Figure size
    % ========================================================================
    width=8.5cm, height=7.5cm,
    %
    % ========================================================================
    % Axis limits recovered from the EPS
    % ========================================================================
    xmin=-16.25, xmax=11.25,
    ymin=-30.5, ymax=4.392,
    %
    % ========================================================================
    % Axis labels
    % ========================================================================
    xlabel={SNR [dB]},
    ylabel={NMSE (dB)},
    xlabel near ticks,
    ylabel near ticks,
    label style={font=\footnotesize},
    %
    % ========================================================================
    % Major ticks
    % ========================================================================
    xtick={-15,-10,...,10},
    ytick={-30,-25,-20,...,0},
    ticklabel style={font=\footnotesize},
    %
    % ========================================================================
    % Legend: lower-left position as in the EPS
    % ========================================================================
    legend style={
        at={(0.98,0.98)},
        %anchor=south west,
        anchor=north east,
        font=\scriptsize,
        inner sep=1pt,
        fill opacity=0.85,
        draw=black!20,
        draw opacity=1,
        text opacity=1
    },
    legend cell align=left,
    %
    % ========================================================================
    % Major grid
    % ========================================================================
    grid=major,
    major grid style={
        draw=black!19,
        line width=0.3pt,
        dash pattern=on 0.3pt off 0.5pt
    },
    mark options={solid}
]

% ============================================================================
% Curve 1: GB-SMCF based GPR, n_t = 16
% ============================================================================
\addplot[
    line width=1pt,
    plotBlue,
    dash pattern=on 6.29pt off 2.72pt,
    mark=o,
    mark size=1.85pt,
    mark options={solid, fill=white, line width=0.6pt}
]
table [x=SNR_dB, y=GB_SMCF_GPR_nt16, col sep=comma]
{Figures/files_txt/nmse_plot_SV_GB_SM_Mat_SE_16.txt};
\addlegendentry{GB-SMCF based GPR ($n_{\textrm{t}}=16$)}

% ============================================================================
% Curve 2: GB-SMCF based GPR, n_t = 8
% ============================================================================
\addplot[
    line width=1pt,
    plotRed,
    solid,
    mark=square,
    mark size=1.85pt,
    mark options={solid, fill=white, line width=0.6pt}
]
table [x=SNR_dB, y=GB_SMCF_GPR_nt8, col sep=comma]
{Figures/files_txt/nmse_plot_SV_GB_SM_Mat_SE_16.txt};
\addlegendentry{GB-SMCF based GPR ($n_{\textrm{t}}=8$)}

% ============================================================================
% Curve 3: GB-SMCF based GPR, n_t = 4
% ============================================================================
\addplot[
    line width=1pt,
    plotGreen,
    dash pattern=on 10.88pt off 2.72pt on 1.70pt off 2.72pt,
    mark=triangle,
    mark size=2.1pt,
    mark options={solid, fill=white, line width=0.6pt}
]
table [x=SNR_dB, y=GB_SMCF_GPR_nt4, col sep=comma]
{Figures/files_txt/nmse_plot_SV_GB_SM_Mat_SE_16.txt};
\addlegendentry{GB-SMCF based GPR ($n_{\textrm{t}}=4$)}

% ============================================================================
% Curve 4: Matern based GPR, n_t = 16
% ============================================================================
\addplot[
    line width=1pt,
    plotPurple,
    dash pattern=on 10.20pt off 3.40pt,
    mark=nmseOpenHexagon,
    mark options={solid, fill=white, line width=0.6pt}
]
table [x=SNR_dB, y=Matern_GPR_nt16, col sep=comma]
{Figures/files_txt/nmse_plot_SV_GB_SM_Mat_SE_16.txt};
\addlegendentry{Mat\'ern based GPR ($n_{\textrm{t}}=16$)}

% ============================================================================
% Curve 5: Squared-exponential (SqExp) based GPR, n_t = 16
% ============================================================================
\addplot[
    line width=1pt,
    plotBrown,
    dash pattern=on 1.70pt off 1.70pt,
    mark=triangle,
    mark size=2.1pt,
    mark options={solid, fill=white, line width=0.6pt, rotate=-90}
]
table [x=SNR_dB, y=SqExp_GPR_nt16, col sep=comma]
{Figures/files_txt/nmse_plot_SV_GB_SM_Mat_SE_16.txt};
\addlegendentry{SqExp based GPR ($n_{\textrm{t}}=16$)}

% ============================================================================
% Curve 6: Least-squares (LS), n_t = 16
% ============================================================================
\addplot[
    line width=1pt,
    black,
    dash pattern=on 6.29pt off 2.72pt,
    mark=diamond,
    mark size=2.55pt,
    mark options={solid, fill=white, line width=0.6pt}
]
table [x=SNR_dB, y=LS_nt16, col sep=comma]
{Figures/files_txt/nmse_plot_SV_GB_SM_Mat_SE_16.txt};
\addlegendentry{LS ($n_{\textrm{t}}=16$)}

\end{axis}

\end{tikzpicture}

%% file: Figures/nmse_plot_CDL_16.tex
% ============================================================================
% Colors 
% ============================================================================
\definecolor{plotBlue}{RGB}{0,114,178}       % GB-SMCF based GPR, n_t = 16
\definecolor{plotGreen}{RGB}{0,158,115}      % GB-SMCF based GPR, n_t = 4
\definecolor{plotOrange}{RGB}{213,94,0}      % LMMSE-isotropic
\definecolor{plotCyan}{RGB}{23,190,207}      % OMP
\definecolor{plotGold}{RGB}{230,159,0}       % AMP

% ============================================================================
% Custom open markers
% ============================================================================
\pgfdeclareplotmark{nmseOpenPlus}{%
    \pgfpathmoveto{\pgfpoint{-0.62pt}{ 1.85pt}}%
    \pgfpathlineto{\pgfpoint{ 0.62pt}{ 1.85pt}}%
    \pgfpathlineto{\pgfpoint{ 0.62pt}{ 0.62pt}}%
    \pgfpathlineto{\pgfpoint{ 1.85pt}{ 0.62pt}}%
    \pgfpathlineto{\pgfpoint{ 1.85pt}{-0.62pt}}%
    \pgfpathlineto{\pgfpoint{ 0.62pt}{-0.62pt}}%
    \pgfpathlineto{\pgfpoint{ 0.62pt}{-1.85pt}}%
    \pgfpathlineto{\pgfpoint{-0.62pt}{-1.85pt}}%
    \pgfpathlineto{\pgfpoint{-0.62pt}{-0.62pt}}%
    \pgfpathlineto{\pgfpoint{-1.85pt}{-0.62pt}}%
    \pgfpathlineto{\pgfpoint{-1.85pt}{ 0.62pt}}%
    \pgfpathlineto{\pgfpoint{-0.62pt}{ 0.62pt}}%
    \pgfpathclose%
    \pgfusepath{fill,stroke}%
}

\pgfdeclareplotmark{nmseOpenX}{%
    \pgfpathmoveto{\pgfpoint{-0.93pt}{ 1.85pt}}%
    \pgfpathlineto{\pgfpoint{ 0.00pt}{ 0.93pt}}%
    \pgfpathlineto{\pgfpoint{ 0.93pt}{ 1.85pt}}%
    \pgfpathlineto{\pgfpoint{ 1.85pt}{ 0.93pt}}%
    \pgfpathlineto{\pgfpoint{ 0.93pt}{ 0.00pt}}%
    \pgfpathlineto{\pgfpoint{ 1.85pt}{-0.93pt}}%
    \pgfpathlineto{\pgfpoint{ 0.93pt}{-1.85pt}}%
    \pgfpathlineto{\pgfpoint{ 0.00pt}{-0.93pt}}%
    \pgfpathlineto{\pgfpoint{-0.93pt}{-1.85pt}}%
    \pgfpathlineto{\pgfpoint{-1.85pt}{-0.93pt}}%
    \pgfpathlineto{\pgfpoint{-0.93pt}{ 0.00pt}}%
    \pgfpathlineto{\pgfpoint{-1.85pt}{ 0.93pt}}%
    \pgfpathclose%
    \pgfusepath{fill,stroke}%
}

\begin{tikzpicture}

\begin{axis}[
    % ========================================================================
    % Figure size
    % ========================================================================
    width=8.5cm, height=7.5cm,
    %
    % ========================================================================
    % Axis limits
    % ========================================================================
    xmin=-16.25, xmax=11.25,
    ymin=-25.735, ymax=6.154,
    %
    % ========================================================================
    % Axis labels
    % ========================================================================
    xlabel={SNR [dB]},
    ylabel={NMSE (dB)},
    xlabel near ticks,
    ylabel near ticks,
    label style={font=\footnotesize},
    %
    % ========================================================================
    % Major ticks
    % ========================================================================
    xtick={-15,-10,...,10},
    ytick={-25,-20,...,5},
    ticklabel style={font=\footnotesize},
    %
    % ========================================================================
    % Legend configuration
    % ========================================================================
    legend style={
        at={(0.98,0.98)},
        anchor=north east,
        font=\scriptsize,
        inner sep=1pt,
        fill opacity=0.85,
        draw=black!20,
        draw opacity=1,
        text opacity=1
    },
    legend cell align=left,
    %
    % ========================================================================
    % Major grid
    % ========================================================================
    grid=major,
    major grid style={
        draw=black!19,
        line width=0.3pt,
        dash pattern=on 0.3pt off 0.5pt
    },
    mark options={solid}
]

% ============================================================================
% Curve 1: GB-SMCF based GPR, n_t = 16
% ============================================================================
\addplot[
    line width=1pt,
    plotBlue,
    dash pattern=on 3.85pt off 1.65pt,
    mark=o,
    mark size=1.85pt,
    mark options={solid, fill=white, line width=0.6pt}
]
table [x=SNR_dB, y=GB_SMCF_GPR_nt16, col sep=comma]
{Figures/files_txt/nmse_plot_CDL_16.txt};
\addlegendentry{GB-SMCF based GPR ($n_{\textrm{t}}=16$)}

% ============================================================================
% Curve 2: GB-SMCF based GPR, n_t = 4
% ============================================================================
\addplot[
    line width=1pt,
    plotGreen,
    dash pattern=on 6.6pt off 1.65pt on 1.05pt off 1.65pt,
    mark=triangle,
    mark size=2.1pt,
    mark options={solid, fill=white, line width=0.6pt}
]
table [x=SNR_dB, y=GB_SMCF_GPR_nt4, col sep=comma]
{Figures/files_txt/nmse_plot_CDL_16.txt};
\addlegendentry{GB-SMCF based GPR ($n_{\textrm{t}}=4$)}

% ============================================================================
% Curve 3: AMP, n_t = 16
% ============================================================================
\addplot[
    line width=1pt,
    plotGold,
    dash pattern=on 3.10pt off 1.05pt on 1.05pt off 1.05pt,
    mark=nmseOpenX,
    mark options={solid, fill=white, line width=0.6pt}
]
table [x=SNR_dB, y=AMP_nt16, col sep=comma]
{Figures/files_txt/nmse_plot_CDL_16.txt};
\addlegendentry{AMP ($n_{\textrm{t}}=16$)}

% ============================================================================
% Curve 4: OMP, n_t = 16
% ============================================================================
\addplot[
    line width=1pt,
    plotCyan,
    dash pattern=on 1.05pt off 1.70pt,
    mark=triangle,
    mark size=2.1pt,
    mark options={solid, fill=white, line width=0.6pt, rotate=180}
]
table [x=SNR_dB, y=OMP_nt16, col sep=comma]
{Figures/files_txt/nmse_plot_CDL_16.txt};
\addlegendentry{OMP ($n_{\textrm{t}}=16$)}

% ============================================================================
% Curve 5: LMMSE-isotropic, n_t = 16
% ============================================================================
\addplot[
    line width=1pt,
    plotOrange,
    dash pattern=on 6.6pt off 1.65pt on 1.05pt off 1.65pt,
    mark=nmseOpenPlus,
    mark options={solid, fill=white, line width=0.6pt}
]
table [x=SNR_dB, y=LMMSE_isotropic_nt16, col sep=comma]
{Figures/files_txt/nmse_plot_CDL_16.txt};
\addlegendentry{LMMSE-isotropic ($n_{\textrm{t}}=16$)}

% ============================================================================
% Curve 6: Least-squares (LS), n_t = 16
% ============================================================================
\addplot[
    line width=1pt,
    black,
    dash pattern=on 3.85pt off 1.65pt,
    mark=diamond,
    mark size=2.55pt,
    mark options={solid, fill=white, line width=0.6pt}
]
table [x=SNR_dB, y=LS_nt16, col sep=comma]
{Figures/files_txt/nmse_plot_CDL_16.txt};
\addlegendentry{LS ($n_{\textrm{t}}=16$)}

\end{axis}

\end{tikzpicture}

%% file: Figures/nmse_plot_SV_CDL_GB_SM_Mat_SE.tex
% ============================================================================

% ============================================================================

% --- Exact RGB colors ------------------------------
\definecolor{plotBlueSVCDL}{RGB}{0,114,178}
\definecolor{plotPurpleSVCDL}{RGB}{127,60,141}
\definecolor{plotBrownSVCDL}{RGB}{140,86,75}

% --- Custom open markers ----------------------------------------------------

\pgfdeclareplotmark{nmseOpenPlus}{%
    \pgfpathmoveto{\pgfpoint{-0.62pt}{ 1.85pt}}%
    \pgfpathlineto{\pgfpoint{ 0.62pt}{ 1.85pt}}%
    \pgfpathlineto{\pgfpoint{ 0.62pt}{ 0.62pt}}%
    \pgfpathlineto{\pgfpoint{ 1.85pt}{ 0.62pt}}%
    \pgfpathlineto{\pgfpoint{ 1.85pt}{-0.62pt}}%
    \pgfpathlineto{\pgfpoint{ 0.62pt}{-0.62pt}}%
    \pgfpathlineto{\pgfpoint{ 0.62pt}{-1.85pt}}%
    \pgfpathlineto{\pgfpoint{-0.62pt}{-1.85pt}}%
    \pgfpathlineto{\pgfpoint{-0.62pt}{-0.62pt}}%
    \pgfpathlineto{\pgfpoint{-1.85pt}{-0.62pt}}%
    \pgfpathlineto{\pgfpoint{-1.85pt}{ 0.62pt}}%
    \pgfpathlineto{\pgfpoint{-0.62pt}{ 0.62pt}}%
    \pgfpathclose%
    \pgfusepath{fill,stroke}%
}

\pgfdeclareplotmark{nmseOpenX}{%
    \pgfpathmoveto{\pgfpoint{-0.93pt}{ 1.85pt}}%
    \pgfpathlineto{\pgfpoint{ 0.00pt}{ 0.93pt}}%
    \pgfpathlineto{\pgfpoint{ 0.93pt}{ 1.85pt}}%
    \pgfpathlineto{\pgfpoint{ 1.85pt}{ 0.93pt}}%
    \pgfpathlineto{\pgfpoint{ 0.93pt}{ 0.00pt}}%
    \pgfpathlineto{\pgfpoint{ 1.85pt}{-0.93pt}}%
    \pgfpathlineto{\pgfpoint{ 0.93pt}{-1.85pt}}%
    \pgfpathlineto{\pgfpoint{ 0.00pt}{-0.93pt}}%
    \pgfpathlineto{\pgfpoint{-0.93pt}{-1.85pt}}%
    \pgfpathlineto{\pgfpoint{-1.85pt}{-0.93pt}}%
    \pgfpathlineto{\pgfpoint{-0.93pt}{ 0.00pt}}%
    \pgfpathlineto{\pgfpoint{-1.85pt}{ 0.93pt}}%
    \pgfpathclose%
    \pgfusepath{fill,stroke}%
}

\begin{tikzpicture}

\begin{axis}[
    % ------------------------------------------------------------------------
    % Figure size
    % ------------------------------------------------------------------------
    width=8.5cm, height=7.5cm,
    % ------------------------------------------------------------------------
    % Axis limits 
    % ------------------------------------------------------------------------
    xmin=-16.25, xmax=11.25,
    ymin=-30.5, ymax=-2.5,
    % ------------------------------------------------------------------------
    % Axis labels
    % ------------------------------------------------------------------------
    xlabel={SNR [dB]},
    ylabel={NMSE (dB)},
    xlabel near ticks,
    ylabel near ticks,
    label style={font=\footnotesize},
    % ------------------------------------------------------------------------
    % Major ticks
    % ------------------------------------------------------------------------
    xtick={-15,-10,...,10},
    ytick={-30,-25,-20,...,0},
    ticklabel style={font=\footnotesize},
    % ------------------------------------------------------------------------
    % Legend: lower-left position 
    % ------------------------------------------------------------------------
    legend style={
        at={(0.01,0.01)},
        anchor=south west,
        font=\scriptsize,
        inner sep=1pt,
        fill=white,
        fill opacity=0.82,
        draw=black!25,
        draw opacity=1,
        text opacity=1
    },
    legend cell align=left,
    % ------------------------------------------------------------------------
    % Grid
    % ------------------------------------------------------------------------
    grid=major,
    major grid style={
        draw=black!19,
        line width=0.3pt,
        dash pattern=on 0.3pt off 0.5pt
    },
    mark options={solid}
]

% ============================================================================
% Curve 1: GB-SMCF based GPR -- SV channel
% ============================================================================
\addplot[
    line width=1pt,
    plotBlueSVCDL,
    dash pattern=on 6.29pt off 2.72pt,
    mark=o,
    mark size=2.0pt,
    mark options={solid, fill=white, line width=0.6pt}
]
table [x=SNR_dB, y=GB_SMCF_GPR_SV, col sep=comma]
{Figures/files_txt/nmse_plot_SV_CDL_GB_SM_Mat_SE.txt};
\addlegendentry{GB-SMCF based GPR SV Channel}

% ============================================================================
% Curve 2: GB-SMCF based GPR -- 3GPP TR38.901 channel
% ============================================================================
\addplot[
    line width=1pt,
    plotBlueSVCDL,
    dash pattern=on 10.88pt off 2.72pt on 1.7pt off 2.72pt,
    mark=nmseOpenPlus,
    mark options={solid, fill=white, line width=0.6pt}
]
table [x=SNR_dB, y=GB_SMCF_GPR_3GPP_TR38901, col sep=comma]
{Figures/files_txt/nmse_plot_SV_CDL_GB_SM_Mat_SE.txt};
\addlegendentry{GB-SMCF based GPR 3GPP TR38.901 Channel}

% ============================================================================
% Curve 3: Matern based GPR -- SV channel
% ============================================================================
\addplot[
    line width=1pt,
    plotPurpleSVCDL,
    dash pattern=on 10.2pt off 3.4pt,
    mark=hexagon,
    mark size=2.15pt,
    mark options={solid, fill=white, line width=0.6pt}
]
table [x=SNR_dB, y=Matern_GPR_SV, col sep=comma]
{Figures/files_txt/nmse_plot_SV_CDL_GB_SM_Mat_SE.txt};
\addlegendentry{Mat\'ern based GPR SV Channel}

% ============================================================================
% Curve 4: Squared-exponential based GPR -- SV channel
% ============================================================================
\addplot[
    line width=1pt,
    plotBrownSVCDL,
    dash pattern=on 1.7pt off 1.7pt,
    mark=triangle,
    mark size=2.1pt,
    mark options={solid, fill=white, line width=0.6pt, rotate=-90}
]
table [x=SNR_dB, y=SqExp_GPR_SV, col sep=comma]
{Figures/files_txt/nmse_plot_SV_CDL_GB_SM_Mat_SE.txt};
\addlegendentry{SqExp based GPR SV Channel}

% ============================================================================
% Curve 5: Matern based GPR -- 3GPP TR38.901 channel
% ============================================================================
\addplot[
    line width=1pt,
    plotBrownSVCDL,
    dash pattern=on 5.1pt off 1.7pt on 1.7pt off 1.7pt,
    mark=nmseOpenX,
    mark options={solid, fill=white, line width=0.6pt}
]
table [x=SNR_dB, y=Matern_GPR_3GPP_TR38901, col sep=comma]
{Figures/files_txt/nmse_plot_SV_CDL_GB_SM_Mat_SE.txt};
\addlegendentry{Mat\'ern based GPR 3GPP TR38.901 Channel}

% ============================================================================
% Curve 6: Squared-exponential based GPR -- 3GPP TR38.901 channel
% ============================================================================
\addplot[
    line width=1pt,
    plotPurpleSVCDL,
    dash pattern=on 1.7pt off 2.805pt,
    mark=triangle,
    mark size=2.1pt,
    mark options={solid, fill=white, line width=0.6pt, rotate=180}
]
table [x=SNR_dB, y=SqExp_GPR_3GPP_TR38901, col sep=comma]
{Figures/files_txt/nmse_plot_SV_CDL_GB_SM_Mat_SE.txt};
\addlegendentry{SqExp based GPR 3GPP TR38.901 Channel}

\end{axis}

\end{tikzpicture}

%% file: Figures/energy_tradeoff.tex
% ============================================================================

% ---------------------------------------------------------------------------
% RGB colors 
% ---------------------------------------------------------------------------
\definecolor{guideBlue}{RGB}{0,114,178}      % n_t = 16 guide line + circle markers
\definecolor{guideRed}{RGB}{214,39,40}       % n_t = 8 guide line + square markers
\definecolor{guideGreen}{RGB}{0,158,115}     % n_t = 4 guide line + triangle markers
\definecolor{guideOrange}{RGB}{230,159,0}    % n_t = 2 guide line

\definecolor{snrCurve01}{RGB}{105,52,118}
\definecolor{snrCurve02}{RGB}{109,83,147}
\definecolor{snrCurve03}{RGB}{101,110,161}
\definecolor{snrCurve04}{RGB}{90,133,165}
\definecolor{snrCurve05}{RGB}{81,155,165}
\definecolor{snrCurve06}{RGB}{76,177,161}
\definecolor{snrCurve07}{RGB}{95,198,147}
\definecolor{snrCurve08}{RGB}{137,215,123}
\definecolor{snrCurve09}{RGB}{196,229,85}
\definecolor{snrCurve10}{RGB}{253,236,81}

\begin{tikzpicture}

% ---------------------------------------------------------------------------
% Main axis: NMSE versus normalized training energy
% ---------------------------------------------------------------------------
\begin{axis}[
    name=mainaxis,
    width=7.5cm,
    height=7.5cm,
    xmode=log,
    log basis x=10,
    xmin=0.1,
    xmax=1.05,
    ymin=-28,
    ymax=1.5,
    xlabel={Normalized training energy},
    ylabel={NMSE [dB]},
    xlabel near ticks,
    ylabel near ticks,
    label style={font=\footnotesize},
    ticklabel style={font=\footnotesize},
    xtick={0.1,0.2,0.3,0.5,1.0},
    xticklabels={0.1,0.2,0.3,0.5,1.0},
    ytick={-30,-25,-20,-15,-10,-5,0,5},
    grid=both,
    major grid style={draw=black!28, line width=0.35pt, dotted},
    minor grid style={draw=black!18, line width=0.25pt, dotted},
    legend style={
        at={(0.02,0.02)},
        anchor=south west,
        font=\scriptsize,
        inner sep=2pt,
        fill opacity=0.9,
        draw=black!30,
        text opacity=1
    },
    legend cell align=left,
    clip mode=individual,
    axis line style={black},
    tick style={black}
]

% ---------------------------------------------------------------------------
% The 10 SNR-colored curves.
% ---------------------------------------------------------------------------
\addplot[line width=1.0pt, snrCurve01, forget plot]
    table [x=train_energy, y=SNR_curve_01, col sep=comma] {Figures/files_txt/energy_tradeoff.txt};
\addplot[line width=1.0pt, snrCurve02, forget plot]
    table [x=train_energy, y=SNR_curve_02, col sep=comma] {Figures/files_txt/energy_tradeoff.txt};
\addplot[line width=1.0pt, snrCurve03, forget plot]
    table [x=train_energy, y=SNR_curve_03, col sep=comma] {Figures/files_txt/energy_tradeoff.txt};
\addplot[line width=1.0pt, snrCurve04, forget plot]
    table [x=train_energy, y=SNR_curve_04, col sep=comma] {Figures/files_txt/energy_tradeoff.txt};
\addplot[line width=1.0pt, snrCurve05, forget plot]
    table [x=train_energy, y=SNR_curve_05, col sep=comma] {Figures/files_txt/energy_tradeoff.txt};
\addplot[line width=1.0pt, snrCurve06, forget plot]
    table [x=train_energy, y=SNR_curve_06, col sep=comma] {Figures/files_txt/energy_tradeoff.txt};
\addplot[line width=1.0pt, snrCurve07, forget plot]
    table [x=train_energy, y=SNR_curve_07, col sep=comma] {Figures/files_txt/energy_tradeoff.txt};
\addplot[line width=1.0pt, snrCurve08, forget plot]
    table [x=train_energy, y=SNR_curve_08, col sep=comma] {Figures/files_txt/energy_tradeoff.txt};

\addplot[line width=1.0pt, snrCurve10, forget plot]
    table [x=train_energy, y=SNR_curve_10, col sep=comma] {Figures/files_txt/energy_tradeoff.txt};

% ---------------------------------------------------------------------------
% Vertical guide lines for the four training-energy cases used in the legend.
% ---------------------------------------------------------------------------
\addplot[line width=1.2pt, guideBlue, dotted]
    coordinates {(1.0,-33) (1.0,5.5)};
\addlegendentry{GB-SMCF based GPR ($n_{\textrm{t}} = 16$)}

\addplot[line width=1.2pt, guideRed]
    coordinates {(0.5,-33) (0.5,5.5)};
\addlegendentry{GB-SMCF based GPR ($n_{\textrm{t}} = 8$)}

\addplot[line width=1.2pt, guideGreen, dash pattern=on 6pt off 1.6pt on 1.1pt off 1.6pt]
    coordinates {(0.25,-33) (0.25,5.5)};
\addlegendentry{GB-SMCF based GPR ($n_{\textrm{t}} = 4$)}

\addplot[line width=1.2pt, guideOrange, dash pattern=on 6pt off 1.6pt on 1.1pt off 1.6pt]
    coordinates {(0.125,-33) (0.125,5.5)};
\addlegendentry{GB-SMCF based GPR ($n_{\textrm{t}} = 2$)}

% ---------------------------------------------------------------------------
% Open markers placed at x = 0.25, 0.5, and 1.0,
% ---------------------------------------------------------------------------
\addplot[
    only marks,
    mark=triangle*,
    mark size=2.4pt,
    guideGreen,
    mark options={solid, fill=white, draw=guideGreen, line width=0.9pt},
    forget plot
] coordinates {(0.25,-5.7511) (0.25,-7.4941) (0.25,-9.2433) (0.25,-10.8915) (0.25,-12.3555) (0.25,-13.6540) (0.25,-14.6452) (0.25,-15.4214) (0.25,-16.2348)};

\addplot[
    only marks,
    mark=square*,
    mark size=2.2pt,
    guideRed,
    mark options={solid, fill=white, draw=guideRed, line width=0.9pt},
    forget plot
] coordinates {(0.5,-8.4676) (0.5,-10.1115) (0.5,-12.3475) (0.5,-13.6522) (0.5,-14.8991) (0.5,-16.7102) (0.5,-18.2104) (0.5,-19.7864) (0.5,-21.9799)};

\addplot[
    only marks,
    mark=*,
    mark size=2.1pt,
    guideBlue,
    mark options={solid, fill=white, draw=guideBlue, line width=0.9pt},
    forget plot
] coordinates {(1.0,-9.8954) (1.0,-11.6274) (1.0,-14.1211) (1.0,-15.6827) (1.0,-17.4011) (1.0,-19.5574) (1.0,-21.5473) (1.0,-23.6198)  (1.0,-27.6151)};

% ---------------------------------------------------------------------------
% Black visual cue on the right-hand side indicating the ordered SNR values.
% ---------------------------------------------------------------------------
\draw[-{Triangle[length=5pt,width=6pt]}, fill=white, line width=0.9pt, black]
    (axis cs:0.82,-26.9) -- (axis cs:0.82,-2.2);
    
\begin{comment}
    \addplot[
    only marks,
    mark=*,
    mark size=1.9pt,
    black,
    mark options={solid, fill=white, draw=black, line width=0.8pt},
    forget plot
] coordinates {(0.82,-9.4386) (0.82,-11.2409) (0.82,-13.6297) (0.82,-14.9633) (0.82,-16.5080) (0.82,-18.5356) (0.82,-20.3704) (0.82,-22.2450) (0.82,-25.9124)};
\end{comment}

% ---------------------------------------------------------------------------
% Rotated SNR annotation with white background
% ---------------------------------------------------------------------------
\node[
    rotate=90,
    font=\footnotesize,
    fill=white,
    draw=black!25,
    line width=0.3pt,
    rounded corners=1pt,
    inner sep=2pt,
    text opacity=1
] at (axis cs:0.71,-15.01)
{SNR [10, 5, 0, -5, -10, -15] dB};

% IMPORTANT: close the MAIN axis here
\end{axis}
% ---------------------------------------------------------------------------
% Separate colorbar axis
% ---------------------------------------------------------------------------
\begin{axis}[
    hide axis,
    scale only axis,
    xmin=0,
    xmax=1,
    ymin=0,
    ymax=1,
    width=0.01pt,
    height=6.00cm,
    at={(mainaxis.north east)},
    anchor=north west,
    xshift=-0.25cm,
    yshift=0.9pt,
    colormap/viridis,
    point meta min=-15,
    point meta max=10,
    colorbar,
    colorbar style={
        width=0.42cm,
        ylabel={SNR [dB]},
        ytick={-15,-10,-5,0,5,10},
        ylabel style={
            font=\footnotesize,
            yshift=4.0mm,
            xshift=3.0mm
        },
        ticklabel style={font=\footnotesize}
    }
]
% Dummy invisible plot used only to generate the colorbar.
\addplot[draw=none] coordinates {(0,0)};
\end{axis}

\end{tikzpicture}

%% file: Figures/se_vs_snr.tex
% ============================================================================

% ============================================================================

% ---------------------------------------------------------------------------
% Exact RGB colors used in the original EPS figure
% ---------------------------------------------------------------------------
\definecolor{sePlotBlue}{RGB}{0,114,178}       % GB-SM GPR, n_t = 16
\definecolor{sePlotRed}{RGB}{214,39,40}        % GB-SM GPR, n_t = 8
\definecolor{sePlotGreen}{RGB}{0,158,115}      % GB-SM GPR, n_t = 4
\definecolor{sePlotOrange}{RGB}{213,94,0}      % LMMSE-isotropic
\definecolor{sePlotCyan}{RGB}{23,190,207}      % OMP
\definecolor{sePerfectGray}{RGB}{51,51,51}     % Perfect CSI

% ---------------------------------------------------------------------------
% Custom open plus marker used by the LMMSE-isotropic curve.
% ---------------------------------------------------------------------------
\pgfdeclareplotmark{seOpenPlus}{%
    \pgfpathmoveto{\pgfpoint{-0.62pt}{ 1.85pt}}%
    \pgfpathlineto{\pgfpoint{ 0.62pt}{ 1.85pt}}%
    \pgfpathlineto{\pgfpoint{ 0.62pt}{ 0.62pt}}%
    \pgfpathlineto{\pgfpoint{ 1.85pt}{ 0.62pt}}%
    \pgfpathlineto{\pgfpoint{ 1.85pt}{-0.62pt}}%
    \pgfpathlineto{\pgfpoint{ 0.62pt}{-0.62pt}}%
    \pgfpathlineto{\pgfpoint{ 0.62pt}{-1.85pt}}%
    \pgfpathlineto{\pgfpoint{-0.62pt}{-1.85pt}}%
    \pgfpathlineto{\pgfpoint{-0.62pt}{-0.62pt}}%
    \pgfpathlineto{\pgfpoint{-1.85pt}{-0.62pt}}%
    \pgfpathlineto{\pgfpoint{-1.85pt}{ 0.62pt}}%
    \pgfpathlineto{\pgfpoint{-0.62pt}{ 0.62pt}}%
    \pgfpathclose%
    \pgfusepath{fill,stroke}%
}

\begin{tikzpicture}

\begin{axis}[
    % -----------------------------------------------------------------------
    % Figure dimensions
    % -----------------------------------------------------------------------
    width=8.5cm,
    height=7.5cm,
    % -----------------------------------------------------------------------
    % Axis limits
    % -----------------------------------------------------------------------
    xmin=-16.25,
    xmax=11.25,
    ymin=0.30,
    ymax=4.20,
    % -----------------------------------------------------------------------
    % Axis labels
    % -----------------------------------------------------------------------
    xlabel={SNR [dB]},
    ylabel={Spectral Efficiency [b/s/Hz]},
    xlabel near ticks,
    ylabel near ticks,
    label style={font=\footnotesize},
    % -----------------------------------------------------------------------
    % Tick locations and fonts
    % -----------------------------------------------------------------------
    xtick={-15,-10,...,10},
    ytick={0.5,1.0,...,4.0},
    ticklabel style={font=\footnotesize},
    % -----------------------------------------------------------------------
    % Legend: upper-left placement
    % -----------------------------------------------------------------------
    legend style={
        at={(0.02,0.98)},
        anchor=north west,
        font=\scriptsize,
        inner sep=1.5pt,
        fill opacity=0.85,
        draw=black!25,
        draw opacity=1,
        text opacity=1
    },
    legend cell align=left,
    % -----------------------------------------------------------------------
    % Grid and marker behavior
    % -----------------------------------------------------------------------
    grid=major,
    major grid style={draw=black!18, line width=0.30pt, dotted},
    mark options={solid}
]

% ============================================================================
% Curve 1: Perfect CSI (no training overhead)
% ============================================================================
\addplot[line width=1.0pt, sePerfectGray, solid]
table [x=SNR_dB, y=Perfect_CSI, col sep=comma]
{Figures/files_txt/se_vs_snr.txt};
\addlegendentry{Perfect CSI (no overhead)}

% ============================================================================
% Curve 2: GB-SM GPR, n_t = 16
% ============================================================================
\addplot[
    line width=1.0pt,
    sePlotBlue,
    dash pattern=on 3.7pt off 1.6pt,
    mark=o,
    mark size=2.2pt,
    mark options={solid, fill=white, draw=sePlotBlue, line width=0.7pt}
]
table [x=SNR_dB, y=GB_SM_GPR_nt16, col sep=comma]
{Figures/files_txt/se_vs_snr.txt};
\addlegendentry{GB-SM GPR ($n_{\textrm{t}}=16$)}

% ============================================================================
% Curve 3: GB-SM GPR, n_t = 8
% ============================================================================
\addplot[
    line width=1.0pt,
    sePlotRed,
    solid,
    mark=square,
    mark size=2.15pt,
    mark options={solid, fill=white, draw=sePlotRed, line width=0.7pt}
]
table [x=SNR_dB, y=GB_SM_GPR_nt8, col sep=comma]
{Figures/files_txt/se_vs_snr.txt};
\addlegendentry{GB-SM GPR ($n_{\textrm{t}}=8$)}

% ============================================================================
% Curve 4: GB-SM GPR, n_t = 4
% ============================================================================
\addplot[
    line width=1.0pt,
    sePlotGreen,
    dash pattern=on 6.4pt off 1.6pt on 1.0pt off 1.6pt,
    mark=triangle,
    mark size=2.35pt,
    mark options={solid, fill=white, draw=sePlotGreen, line width=0.7pt}
]
table [x=SNR_dB, y=GB_SM_GPR_nt4, col sep=comma]
{Figures/files_txt/se_vs_snr.txt};
\addlegendentry{GB-SM GPR ($n_{\textrm{t}}=4$)}

% ============================================================================
% Curve 5: OMP, n_t = 16
% ============================================================================
\addplot[
    line width=1.0pt,
    sePlotCyan,
    dash pattern=on 1.0pt off 1.65pt,
    mark=triangle,
    mark size=2.35pt,
    mark options={solid, fill=white, draw=sePlotCyan, line width=0.7pt, rotate=180}
]
table [x=SNR_dB, y=OMP_nt16, col sep=comma]
{Figures/files_txt/se_vs_snr.txt};
\addlegendentry{OMP ($n_{\textrm{t}}=16$)}

% ============================================================================
% Curve 6: LMMSE-isotropic, n_t = 16
% ============================================================================
\addplot[
    line width=1.0pt,
    sePlotOrange,
    dash pattern=on 6.4pt off 1.6pt on 1.0pt off 1.6pt,
    mark=seOpenPlus,
    mark options={solid, fill=white, draw=sePlotOrange, line width=0.7pt}
]
table [x=SNR_dB, y=LMMSE_isotropic_nt16, col sep=comma]
{Figures/files_txt/se_vs_snr.txt};
\addlegendentry{LMMSE-isotropic ($n_{\textrm{t}}=16$)}

% ============================================================================
% Curve 7: Least-squares, n_t = 16
% ============================================================================
\addplot[
    line width=1.0pt,
    black,
    dash pattern=on 3.7pt off 1.6pt,
    mark=diamond,
    mark size=2.45pt,
    mark options={solid, fill=white, draw=black, line width=0.7pt}
]
table [x=SNR_dB, y=LS_nt16, col sep=comma]
{Figures/files_txt/se_vs_snr.txt};
\addlegendentry{LS ($n_{\textrm{t}}=16$)}

\end{axis}

\end{tikzpicture}